\documentclass[12pt]{article}

\usepackage{classeJBArxive}
\usepackage{amssymb,amsmath,latexsym,array,verbatim}
\usepackage{nomencl}
\usepackage{makecell}
\makenomenclature
\usepackage{graphicx}
\usepackage{verbatim} 
\usepackage{mathtools}
\usepackage{wrapfig,lipsum}
\usepackage{booktabs}
\usepackage[colorlinks]{hyperref}
\usepackage{caption}
\usepackage{diagbox}

\graphicspath {{./figures/}}

\usepackage[font={small,it}]{caption}

\let\norm\relax
\DeclarePairedDelimiterX{\norm}[1]{\lVert}{\rVert}{#1}

\newcommand{\DF}{\textsc{Dufort}--\textsc{Frankel}}
\definecolor{orcidlogocol}{HTML}{A6CE39}
\definecolor{darkviolet}{rgb}{0.31, 0, 0.51}

\title{\vspace{-1.5cm}
Searching an optimal experiment observation sequence \\	
to estimate the thermal properties of a multilayer wall \\
under real climate conditions
\vspace{4pt}}
\author{Ainagul Jumabekova\textsuperscript{a,c}$^{\ast}$, Julien Berger\textsuperscript{b}, \\
 {Aurélie Foucquier\textsuperscript{c}}, George S. Dulikravich\textsuperscript{d} }

\begin{document}

\maketitle

\begin{center}
\small
\textsuperscript{a} Univ. Grenoble Alpes, Univ. Savoie Mont Blanc, UMR 5271 CNRS, LOCIE, 73000 Chambéry, France,  \\
\textsuperscript{b} LaSIE, La Rochelle University, CNRS, UMR 7356, 17000 La Rochelle, France,\\ 
\textsuperscript{c} Univ. Grenoble Alpes, CEA, LITEN, DTS, INES, F-38000, Grenoble, France,\\
\textsuperscript{d} Department of Mechanical and Materials Engineering, Florida International University, Miami, Florida,  \\
$^{\ast}$corresponding author, e-mail address : ainagul.jumabekova@univ-smb.fr,\\ ORCiD : 0000-0001-5554-4249\\

\end{center}
\begin{abstract}
The \emph{in situ} estimation of the thermal properties of existing building wall materials is a computationally expensive procedure. Its cost is highly proportional to the duration of measurements. To decrease the computational cost a methodology using a D-optimum criterion to select an optimal experiment duration is proposed. This criterion allows to accurately estimate the thermal properties of the wall using a reduced measurement plan. The methodology is applied to estimate the thermal conductivity of the three-layer wall of a historical building in France. Three different experiment sequences (one, three and seven days) and three spatial distributions of the thermal conductivity are investigated. Then using the optimal duration of observations the thermal conductivity is estimated using the hybrid optimization method. Results show a significant reduction of computational time; and reliable simulation of physical phenomena using the estimated values.

parameter estimation problem ; hybrid optimization method.

\end{abstract}

\section{Introduction}

Recently, reducing carbon emissions has become one of the most important tasks worldwide. The building sector is responsible for approximately one-third of global energy-related carbon emissions~\cite{IEA_2019}. Implementation of energy-efficient policies can significantly decrease these emissions. Policies should focus on minimizing energy demand for heating and cooling through the retrofitting of the existing building stock. The success of the retrofitting strategies highly depends on accurate predictions of the building's energy performance. Building simulation programs play the main role in the efficiency evaluation of energy-saving policies. However, to obtain reliable prediction simulation software requires accurate values of the building material properties. As recent studies show, there is a gap between calculated and actual energy consumption~\cite{Sunikka_2012,Wingfield_2008}. Specifically, this discrepancy comes from the uncertainty of the thermophysical characteristics of the building wall. Thermal conductivity and heat capacity can be inferred through the solution of the so-called inverse problem. The latter corresponds to an optimization problem, which aims to minimize a difference between direct model outputs and experimental observations. 

The acquisition of experimental data in existing buildings faces several constraints. First, dealing with existing buildings should consider their residents, and not interfere in occupants' everyday life. The next factor is the cost of experimental design. For instance, the quality of the experiment can be improved by installing more sensors, which would increase the cost of the experiment. Finally, the duration of an experiment should be questioned. On the one hand, longer experiments guarantee better accuracy of estimated parameters. But on the other hand, the computational cost of the parameter estimation problem is highly proportional to the duration of the observations. Therefore, it is important to address this issue and find an optimum between the "richness" of the experimental data and the computational cost of the inverse problem. 

Several studies that deal with this dilemma can be found in the literature. The first approach presents an error measurement of thermal conductivity and heat capacity calculated over a different number of days, from one up to twenty~\cite{RODLER_2019}. Through a comparison of the relative errors and parameters dispersion for each period, the optimal number of days is chosen. However, this approach is not reliable for a longer experiment duration, for instance several months, due to computational cost. Another drawback is a lack of reliable criteria, so the choice is made based on a personal perspective. A second practice is maximizing the so-called D-optimum criterion, the determinant of the sensitivity matrix and its transpose, which minimizes the confidence regions of the parameters. Several articles apply this criterion on mass transfer in a porous building material~\cite{BERGER_2017_OED, Berger_2019}. Studies of optimal experiment design of heat transfer have been performed in controlled laboratory conditions. For instance, in~\cite{DALESSANDRO_2019}, the optimal heating period and the duration of the experiment were investigated for a three-layer experimental set-up, where a thin heater is placed between two identical samples. Unfortunately, the aforementioned articles do not apply the optimal experiment design in real climate conditions. 

This article presents a real case study of the wall of a historical building in France. Temperature measurements were taken over one year both on the wall surfaces and within the wall using five different sensors. The wall consists of three layers. Previously, the thermal conductivity of the wall were identified by implementing the Bayesian approach~\cite{BERGER_2016}. However, an order of $10^{\,5}$ direct model computations were required to solve the estimation problem using the whole set of observations. Therefore, it is crucial to decrease the measurement period and to preserve the accuracy of the estimated parameters at the same time.

The aim of the article is to propose a methodology to choose the optimal experiment duration for the estimation of thermal conductivity.
By using the D-optimum criterion~\cite{beck_1977} and the advantages of the~\DF\ numerical scheme, this approach can be used to choose the best period of the experiment efficiently. Additionally, thermal conductivity is estimated over the chosen period using the hybrid optimization method. Hybrid optimization methods combine gradient and heuristic optimization strategies to find rapidly global extremum.

The article is organized as follows: Section~\ref{sec:methodology} presents the mathematical and numerical models together with the methodology to solve the parameter estimation problem. Section~\ref{sec:case_study} introduces a case study of a multi-layer wall. The estimation of the thermal conductivity of each layer of the wall requires several steps. First, the identifiability of the parameters is demonstrated. Next, the optimal duration of the experiment is chosen. Then, the results of the parameter estimation problem are given. Finally, the reliability of a whole approach is discussed. 

\section{Methodology}
\label{sec:methodology}
\subsection{Physical Model}

\begin{figure}[!ht] 
\centering
\includegraphics[width=0.5\linewidth]{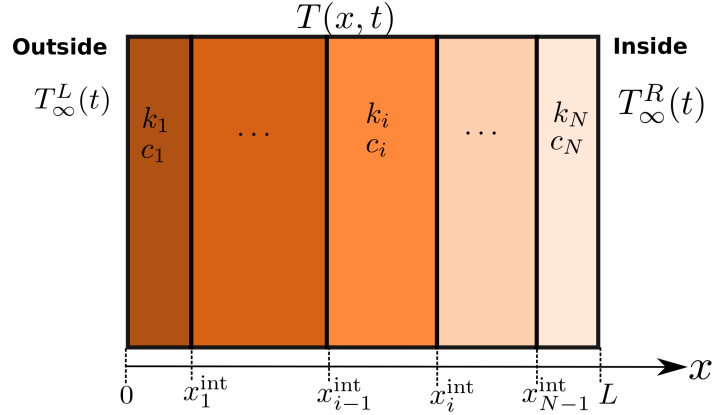}
\caption{Illustration of the wall construction.}
\label{fig:wall_1}
\end{figure}

The physical problem considers one--dimensional heat conduction transfer through a building wall. The wall is composed of $N$ layers, each layer differs from the other by its thermal properties and thickness, as shown in Figure~\ref{fig:wall_1}. The temperature in the wall is defined on the domains  $\Omega_{\,x}\,:\, x \in [\,0\,,L\,]\,$ and $ \Omega_{\,t}\,:\,t \in [\,0\,,\tau_{\,\mathrm{max}}\,]\,$, where $L\,\mathsf{[\,m\,]}$ is the length of the wall and $\tau_{\,\mathrm{max}} \,\mathsf{[\,s\,]}$ is the duration of the experiment:
\begin{align*}
T\,:\, [\,0\,,L\,]\, \times \,[\,0\,,\tau_{\,\mathrm{max}}\,] \, \longrightarrow \,\mathbb{R}\,.
\end{align*}
The mathematical formulation of the heat transfer process is given below: 
\begin{align}
\label{eq:heat_eq_ph}
c\;\pd{T}{t} \egal \pd{}{x}\,\Bigl(\,k\;\pd{T}{x}\,\Bigr)\,,
\end{align}
where $c\,\mathsf{[\,J\cdot K^{-1} \cdot m^{-3}\,]}$ is the volumetric heat capacity, or $c \egal \rho \,\cdot \, c_{\,p}\,$, corresponding to the product between the material density $\rho\,\mathsf{[\,kg \cdot m^{-3}\,]}$  and the specific heat $c_{\,p}\,\mathsf{[\,J \cdot kg^{-1} \cdot K^{-1}\,]}$, and $k\,\mathsf{[\,W \cdot m^{-1} \cdot K^{-1}\,]}$ is the thermal conductivity. Both properties depend on the space coordinate. \\*
The inside and outside surface temperatures of the wall are set as boundary conditions:
\begin{align}
T \egal & T_{\,\infty}^{\,L}\,\bigl(\,t\,\bigr)\,, \qquad x\egal 0\,,\\ 
T \egal & T_{\,\infty}^{\,R}\,\bigl(\,t\,\bigr)\,, \qquad x\egal L\,.
\end{align}
The initial condition of the problem is defined as a solution of the steady state problem:
\begin{align}
T \egal & T_{\,0}\,\bigl(\,x\,\bigr)\,, \qquad t\egal 0\,.
\end{align}
The contact between layers is assumed to be perfect,  thereby imposing continuity on the temperature and the heat flux~\cite{stephenson1971,Ozisik}.
\begin{align}
 \Bigl[\, T\,\bigl(\,x\,,\,t\,\bigr)\, \Bigr]_{\,x\egal x^{\,\mathrm{int}}_{\,i}} \egal  0\,,\qquad  \Bigl[\, k\;\pd{T\,\bigl(\,x\,,\,t\,\bigr)}{x}\, \Bigr]_{\,x\egal x^{\,\mathrm{int}}_{\,i}} \egal  0\,.
\end{align}

\subsection{Modelling the thermophysical properties} 
\label{sec:model_properties}
This section outlines a mathematical model to represent space dependant  thermal conductivity $k$ and volumetric heat capacity $c$. Generally, the spatial distribution of the thermophysical properties is defined as a \textbf{piecewise function}.
The expression for thermal conductivity $k$ is formulated as: 
\begin{align}
k\,(\,x\,) \egal \sum_{i=1}^N\,k_{\,i}\;\varphi_{\,i}\,(\,x\,)\,,
\end{align} 
while volumetric heat capacity $c$ is defined as: 
\begin{align}
c\,(\,x\,)  \egal \sum_{i=1}^N\,c_{\,i}\;\varphi_{\,i}\,(\,x\,)\,,
\end{align} 
where $\bigl\{\,\varphi_{\,i}\,(\,x\,)\,\bigr\}_{\,i \egal 1\,,\ldots\,,N\, }$ are piecewise functions and can be written as:
\begin{align}
\varphi_{\,i}\,(\,x\,) \egal \begin{cases}
1 \,, & x^{\,\mathrm{int}}_{\,i-1}\leqslant x \leqslant x^{\,\mathrm{int}}_{\,i} \,, \quad i \egal {1\,,\ldots\,,N}\,, \\
0 \,, & \mathrm{otherwise}\,,
\end{cases} \,
\end{align}
where $x^{\,\mathrm{int}}_{\,i-1}$ and $x^{\,\mathrm{int}}_{\,i}$ are the left and right interfaces of the layer $i$, respectively.

This article focuses solely on the estimation of the thermal conductivity of the wall. Therefore, the piecewise parameterization is used to characterize the variation of the volumetric heat capacity. However, thermal conductivity may depend on moisture content in a wall that changes spatially due to daily cycles of temperature and relative humidity~\cite{Rode_2004}, therefore there are several options to define the spatial variation of the thermal conductivity. It can be presented as a linear combination of basis functions:
\begin{align}
k\,(\,x\,) \egal \sum_{i=1}^M\,\beta_{\,i}\,Y_{\,i}\,(\,x\,)\,,
\end{align}
where $M$ is number of the chosen basis functions $\{\,Y_{\,i}\,(\,x\,)\,\}$.   \\*
This article deals with three different types of the properties parameterization : the standard--case scenario as piecewise functions, the polynomial interpolation, the spline interpolation. The first is defined above. The next parameterization is found through \textbf{a polynomial interpolation}.
Thermal conductivity is formulated by standard series of the polynomial basis functions, and it yields to:
\begin{align}
k\,(\,x\,) \egal k_{\,0} \plus \beta_{\,1}\,x \plus \beta_{\,2}\,x^{\,2} \plus \ldots \plus \beta_{\,M}\,x^{\,M} \,, \qquad 0 \leqslant x \leqslant L\,.
\end{align} 
Finally, thermal conductivity is presented using \textbf{a piecewise polynomial interpolation}. Therefore, 
\begin{align}
k\,(\,x\,) \egal Y_{\,i}\,(\,x\,)\,, \qquad x^{\,\mathrm{int}}_{\,i-1}\leqslant x \leqslant x^{\,\mathrm{int}}_{\,i} \,,
\end{align}
where $Y_{\,i}\,(\,x\,)$ is polynomial with a different degree for each wall layer, or
\begin{align}
Y_{\,i}\,(\,x\,) \egal \sum_{r=0}^G\,\beta_{\,r}\,x^{\,r}\,,
\end{align}
where $G$ is the chosen polynomial order.

The further choice of a parameterization will depend on the structural identifiability of unknown parameters since it varies due to number of available measurements.

\subsection{Dimensionless equation}

This section introduces the dimensionless model equations. 

Let us define the following dimensionless variables:
\begin{align}
x^{\,\star} & \egal \dfrac{x}{L}\,, & u & \egal   \dfrac{T \moins T_{\,\mathrm{ref}}}{\Delta T_{\,\mathrm{ref}}}\,,  & t^{\,\star} & \egal  \dfrac{t}{t_{\,\mathrm{ref}}}\,,\\
 k^{\,\star} & \egal  \dfrac{k}{k_{\,\mathrm{ref}}}\,,  & c^{\,\star} & \egal \dfrac{c}{c_{\,\mathrm{ref}}}\,,  & \mathrm{Fo} & \egal  \dfrac{t_{\,\mathrm{ref}}\,\cdot\,k_{\,\mathrm{ref}}}{L^{\,2}\,\cdot\,c_{\,\mathrm{ref}}}\,, \nonumber
\end{align}
where subscripts $\boldsymbol{\mathrm{ref}}$ relate to a characteristic reference value, and superscript $\boldsymbol{\star}$ for dimensionless parameters. Thus, equation~\eqref{eq:heat_eq_ph} becomes:
\begin{align} \label{eq:dim_heat_eq}
c^{\,\star}\,(\,x^{\,\star}\,)\,\pd{u}{t^{\,\star}}  \egal \mathrm{Fo}\,\pd{}{x^{\,\star}}\,\Bigl(\,k^{\,\star}\,(\,x^{\,\star}\,)\,\pd{u}{x^{\,\star}}\,\Bigr)\,.
\end{align}

{Dirichlet}--type boundary conditions are converted to:
\begin{align}
u \egal & u_{\,\mathrm{L}}\,(\,t^{\,\star}\,)\,, \qquad x^{\,\star}\egal 0\,,\qquad \mathrm{where}\qquad u_{\,\mathrm{L}} \egal \dfrac{T_{\,\infty}^{\,L} \moins T_{\,\mathrm{ref}}}{\Delta T_{\,\mathrm{ref}}}\,,\\ 
u \egal & u_{\,\mathrm{R}}\,(\,t^{\,\star}\,)\,, \qquad x^{\,\star}\egal 1\,,\qquad \mathrm{where}\qquad u_{\,\mathrm{R}} \egal \dfrac{T_{\,\infty}^{\,R}\moins T_{\,\mathrm{ref}}}{\Delta T_{\,\mathrm{ref}}}\,.
\end{align}

The initial condition is transformed to:
\begin{align}
u \egal u_{\,0}\,(\,x^{\,\star}\,)\,,\qquad \mathrm{where}\qquad u_{\,0} \egal \dfrac{T_{\,0} \moins T_{\,\mathrm{ref}}}{\Delta T_{\,\mathrm{ref}}}\,.
\end{align}

\subsection{Numerical Model}
\label{sec:num_mod}
After defining the governing equation, this section details the construction of the numerical model.  Let us discretize uniformly  the space and time intervals, with the parameters $\Delta x^{\star}$ and $\Delta t^{\star}$, respectively. The discrete values of function $u\,(\,x^{\star}\,,t^{\star}\,)$ are defined as $\,u_{\,j}^{\,n} \ \eqdef \ u\,(\,x^{\star}_{\,j}\,,t^{\star}_{\,n}\,)\,$, where $\,j \, \in \, \{\,1,\,\ldots\,,N_{\,x}\, \}\, $ and $\,n\, \in \, \{\,1,\,\ldots\,,N_t\, \} $. The solution  $u\,(\,x^{\star}\,,t^{\star}\,)$ is obtained using the  {Dufort}-{Frankel} numerical scheme. This numerical model is chosen due to its explicit formulation without loss of accuracy or reliability~\cite{Du_Fort_1953,Gasparin_2017a,Gasparin_2017b}.

For the non-linear case, the solution is calculated with the following expression:
\begin{align}
\label{eq:num_solu} 
u_{\,j}^{\,n+1} \egal \nu_{\,1}\cdot u_{\,j+1}^{\,n} \plus \nu_{\,2}\cdot u_{\,j-1}^{\,n} \plus \nu_{\,3}\cdot u_{\,j}^{\,n-1}\,, 
\end{align}
where
\begin{align}
&\nu_{\,1} \egal \frac{\lambda_{\,1}}{\lambda_{\,0} \plus \lambda_{\,3}}\,,
&&\nu_{\,2} \egal \frac{\lambda_{\,2}}{\lambda_{\,0} \plus \lambda_{\,3}}\,,
&&&\nu_{\,3} \egal \frac{\lambda_{\,0} \moins \lambda_{\,3}}{\lambda_{\,0} \plus \lambda_{\,3}}\,,
\end{align}
and
\begin{align}
&\lambda_{\,0} \ \eqdef \ 1\,, 
&&\lambda_{\,3} \ \eqdef \ \frac{\Delta t^{\star}}{\Delta x^{\star\,2}}\,\frac{\mathrm{Fo}}{c^{\star}_{\,j}}\,\bigl(\,k^{\star}_{j\plus\frac{1}{2}}\plus k^{\star}_{j\moins\frac{1}{2}}\,\bigr)\,\\
&\lambda_{\,1} \ \eqdef \ \frac{2\,\Delta t^{\star}}{\Delta x^{\star\,2}}\,\frac{\mathrm{Fo}}{c^{\star}_{\,j}}\,k^{\star}_{j\plus\frac{1}{2}}\,, 
&& \lambda_{\,2} \ \eqdef \ \frac{2\,\Delta t^{\star}}{\Delta x^{\star\,2}}\,\frac{\mathrm{Fo}}{c^{\star}_{\,j}}\,k^{\star}_{j\moins\frac{1}{2}}\,. 
\end{align}
The nonlinear coefficients are approximated by:
\begin{align}
k^{\star}_{j\,\pm\,\frac{1}{2}} \egal k^{\star}\,\biggl(\,\frac{x^{\star}_{\,j}\,+\,x^{\star}_{\,j\,\pm\,1}}{2}\,\biggr)\,.
\end{align}
\subsection{Parameter Estimation Problem}
The next section presents different steps required to solve the parameter estimation problem. The issue is to determine the thermal conductivity within each layer of the wall. Each case of the thermal conductivity parameterization holds its own set $\mathrm{P}$ of unknown dimensionless parameters. \\*
The parameter set of the \textbf{piecewise representation} directly corresponds to the thermal conductivity values of wall layers:  
\begin{align}
\mathrm{P} \egal  \{\,k^{\,\star}_{\,1}\,,k^{\,\star}_{\,2}\,,\,\ldots\,\,,k^{\,\star}_{\,N}\,\} \,.
\end{align}
The set of the \textbf{polynomial parameterization} consists of polynomial coefficients, and its number depends on the chosen order of the series:
\begin{align}
\mathrm{P} \egal  \{\,k^{\,\star}_{\,0}\,,\beta_{\,1}\,,\beta_{\,2}\,,\,\ldots\,\,,\beta_{\,M}\,\} \,.
\end{align}
Unlike the polynomial representation, the number of the coefficients of  
the \textbf{spline interpolation} differs depending on the layer. Therefore, its parameters set is expressed as:
\begin{align}
\mathrm{P} \egal  \displaystyle {\bigcup\limits_{i\egal 1}^{\,N}\, \{\,\beta_{\,i}^{\,0}\,,\,\beta_{\,i}^{\,1}\,,\,\ldots\,\,,\beta_{\,i}^{\,G}\,\}} \,.
\end{align}
We define the component of the parameter set $\mathrm{P}$ as $\mathrm{P}_{\,m} \, \in \ \mathrm{P}$ with $\,m\, \in \, \{\,1,\,\ldots\,,N_{\,p}\, \} \,$, where $N_{\,p}$ is the total number of unknown parameters and varies depending of chosen parameterization. In case of piecewise it equals to the number of all wall's layers, or $N_{\,p} \egal N $; for polynomial and spline interpolations  $N_{\,p} \egal M $ and $N_{\,p} \egal G \,\times\, N$ respectively. Therefore, the aim is to estimate these parameters  $\mathrm{P}_{\,m}$. \\*
The recovery of the parameters $\mathrm{P}$ is based on the minimization of a difference between the computed temperature distribution $u\,(\,x\,,t\,)$ and the given temperature measurements $u^{\,\mathrm{obs}}$. The cost of this minimization procedure is highly proportional to the number of parameters and the length of the observations. Thus, it is important to check whether all parameters can be identified, and if the estimation process requires all the measurement data. The next sections introduce a methodology on how to answer these questions.

\subsubsection{Structural Identifiability}
Initially, it is of capital importance to check whether the parameters can be identified independently from the measurements. One may introduce the formal identifiability of the unknown parameters as follows. We assume here  $u\,(\,x\,,t\,)$ is the only observable field.

A parameter $\mathrm{P}_{\,m} \,\in\, \mathrm{P}$ is Structurally Globally Identifiable (SGI) if the following condition is satisfied~\cite{Walter_1982}:
\begin{align}
\forall t \,, \qquad u\,(\,\mathrm{P}\,) \egal u\,(\,\mathrm{P}^{\,'}\,) \,\Rightarrow \,\mathrm{P}_{\,m} \egal \mathrm{P}^{\,'}_{\,m} \,.
\end{align}
This property should be demonstrated for each type of parametrization.
\subsubsection{Practical Identifiability}
Before performing the parameter identification process we need to study whether the parameters can be estimated regarding the given experimental design.
These results are obtained using practical identifiability by calculating the sensitivity coefficients. 
The sensitivity coefficient is defined as the first derivative of the (numerical) observations with respect to an unknown parameter \cite{Finsterle_2015,Walter_1990}:
\begin{align}
X_{\,\mathrm{P}_{\,m}}  \egal \pd{\mathrm{u}}{\mathrm{P}_{\,m}} \,.
\end{align}

The function $X_{\,\mathrm{P}_{\,m}}$ measures the sensitivity of the estimated field $u$ with respect to changes in the parameter $\mathrm{P}_{\,m}$. If the value of $X_{\,\mathrm{P}_{\,m}}$ is small it can be concluded that that the parameter $\mathrm{P}_{\,m}$ does not influence the output $u$ and cannot be identified with accuracy. Moreover, the inverse problem is also ill--conditioned. If the sensitivity coefficients are linearly dependent, the inverse problem becomes ill--posed. Therefore, to get the best estimation of parameters $\mathrm{P}$, it is
necessary to have linearly--independent sensitivity functions $X_{\,\mathrm{P}_{\,m}}$ with large magnitudes for all the parameters $\mathrm{P}_{\,m}$. 

The sensitivity coefficients are computed by direct differentiation of the governing equation~\eqref{eq:dim_heat_eq} with respect to an unknown parameter.
Each parameter set implies a different set of sensitivity equations.
For instance, if the thermal conductivity is formulated as a piecewise function, the sensitivity coefficient of the unknown parameter $k_{\,i}^{\,\star}$ is given by the following differential equation:
\begin{align}
\pd{X_{\,k^{\,\star}_{\,i}}}{t^{\,\star}} = \frac{\mathrm{Fo}}{c^{\,\star}}\,\pd{}{x^{\,\star}}\,\biggl(\,\pd{k^{\,\star}}{k^{\,\star}_{\,i}}\,\pd{u}{x^{\,\star}} + \,k^{\,\star}\,\pd{X_{\,k^{\,\star}_{\,i}}}{x^{\,\star}}\,\biggr)\,,
\end{align}
where $X_{\,k_{\,i}^{\,\star}}\, \egal \, \displaystyle \pd{u}{k_{\,i}^{\,\star}}$, and the derivative of the thermal conductivity yields to:
\begin{align}
    \pd{k^{\,\star}}{k^{\,\star}_{\,i}} \egal \begin{cases}
    1\,,\, x^{\,\mathrm{int}}_{\,i-1} \leqslant x \leqslant x^{\,\mathrm{int}}_{\,i} \,,\\
    0\,, \,\text{otherwise}\,.
    \end{cases}
\end{align}

Similarly, sensitivity equations for coefficients of the polynomial and spline representations are retrieved. The difference lies in the calculation of the thermal conductivity derivative.
In case of polynomial parameterization, the derivative is computed as: 
\begin{align}
    \pd{k^{\,\star}}{k^{\,\star}_{\,0}} \egal 1\,,\qquad \text{or} \qquad \pd{k^{\,\star}}{\beta_{\,i}} \egal x^{\,i-1}\,.
\end{align}
An analogous expression is obtained for the spline parameterization: 
\begin{align}
    \pd{k^{\,\star}}{\beta_{\,r}} \egal x^{\,r}\,, \qquad r\egal \{\,0\,,\,\ldots\,, G\,\}\,,\qquad  \, x^{\,\mathrm{int}}_{\,i-1} \leqslant x \leqslant x^{\,\mathrm{int}}_{\,i} \,.
\end{align}
 
\subsubsection{The Optimal Experiment Design}

Finally, to improve the precision of the estimated results, it is crucial to find an optimal experiment design (OED). Under the optimal experiment, we mean measurement conditions that maximize the estimated parameter accuracy. The purpose of this study is to determine the optimum duration of experiment $\delta\,\tau$, to retrieve which period of observations provide us with maximum accuracy. To search for this optimal experiment design, we introduce the following measurement plan:
\begin{align}
\pi \egal \{\,\delta\,\tau\,\}\,.
\end{align}
The choice of the experimental design relies on the maximization of certain quality indicators~\cite{Ucinski}, thus the objective of OED is to find a measurement plan $\hat{\pi}$ that maximizes a chosen function $\Psi$:
\begin{align}
\hat{\pi}\egal & \underset{\pi}{\mathrm{argmax}}\,\Psi\,\,.
\end{align}
Several objective functions $\Psi$ can be applied, in this article a measurement plan is analyzed using a D--optimum criterion:
\begin{align}
\Psi \egal \det\,\widetilde{\mathrm{F}}\,(\,\pi\,)\,,
\end{align}
where $\widetilde{\mathrm{F}}\,(\,\pi\,)$ is the modified \textsc{Fisher} matrix. The elements of the matrix represent the average value of the parameters sensitivity during the measurement plan $\,\pi\,$ and are defined according to~\cite{beck_1977}:
\begin{align}
\label{eq:fisher}
\widetilde{\mathrm{F}}\,(\,\pi\,) & \egal \Bigl[\,\widetilde{\mathrm{F}}_{\,ij\,}\,\Bigr], \qquad \forall \,\bigl(\,i,\,j\,\bigr) \, \in \, \{ \,1,...,N_p\,\} \,, \\
\widetilde{\mathrm{F}}_{\,ij\,} \egal & \displaystyle \nicefrac{1}{\sigma^{\,2}}\, \displaystyle \,\sum_{q\egal 1}^{N_{\,\text{m}}}\,\frac{1}{\delta \tau}\,\int_{\,t_{\,\mathrm{ini}}}^{\,t_{\,\mathrm{ini}} \plus \delta \tau}{ X_{\,P_{\,i}} \, X_{\,P_{\,j}} \; \mathrm{dt}}\,,
\end{align}
where $X_{\,P_{\,i}}$ is the sensitivity coefficient of the solution related to the parameter $\mathrm{P}_{\,i}$, $\sigma$ is the measurement uncertainty, $N_{\,p}$ is the number of parameters, and $N_{\,\text{m}}$ is the total number of measurements, and $\delta\,\tau$ is the duration of experiment, which should be investigated. Additionally, the initial time of measurements $t_{\,\mathrm{ini}} \,\in\, [\,0\,,\tau_{\,\mathrm{max}}\,]$ and the total number of periods  $N_{\,\delta\,\tau}$  in the whole interval $[\,0\,,\tau_{\,\mathrm{max}}\,]$ are defined. 

For example, to find an optimal week during one year of observations fifty two \textsc{Fisher} matrices should be investigated, therefore, in this case  $\delta\,\tau \egal 7\,[\,\mathsf{days}\,]$, $t_{\,\mathrm{ini}}$ shifts every seven days from the first observation day and $N_{\,\delta\,\tau} \egal 52\,$.  

Thus, from a numerical point of view, to detect the optimal period of measurements one should choose the duration of the experiment, calculate the modified \textsc{Fisher} matrix for the chosen measurement plan and then find an optimal sequence that maximizes the D--optimum criterion.

\subsubsection{Cost function minimization}
\label{sec:cost_min}
After finding the optimal sequence, the parameter estimation problem is solved by minimizing the following cost function by the optimization method:
\begin{equation}
\mathrm{J}\,\bigl(\,\mathrm{P}\,\bigr)
\egal \sum_{\,i\egal 1}^{\,M} \omega_{\,i}\,\Biggl|\Biggl|\,{u^{\,\mathrm{num}}\,\bigl(\,x^{\,\star}\egal x^{\,\star}_{\,i}\,,\,\mathrm{P}\,\bigr) \moins u_{\,i}^{\,\mathrm{obs}}}\,\Biggr|\Biggr|_{\,2} 
\end{equation}
The value of $u^{\,\mathrm{num}}$ results from the solution of the direct problem~\eqref{eq:dim_heat_eq} for a given set of parameters $\mathrm{P}$. The values of $u_{\,i}^{\,\mathrm{obs}}$ are given by the measurements at the points $x^{\,\star}\egal x_{\,i}^{\,\star}$ respectively, $M$ is the total number of sensors. It is assumed that measurement errors are additive with zero mean, constant variance, uncorrelated and normal distribution. The weights are calculated as $\omega_{\,i} \egal \dfrac{1}{\sigma_{i}^{\,2}}$, where $\sigma_{\,i}$ is a standard deviation of the measurement $u_{\,i}^{\,\mathrm{obs}}$ of the $i-$th sensor \cite{Ozisik}. The norm is calculated according to:
\begin{align}
\Bigl|\Bigl|\,y\,\Bigr|\Bigr|_{\,2} \egal \int_{\,0}^{t_{\,\mathrm{max}}} \, \Bigl(\,y\,(\,t\,)\,\Bigr)^{\,2} \,\mathrm{d}t
\end{align}
\emph{A priori} parameters values, denoted as $\mathrm{P^{\,\circ}}$, are used in the minimization process and to compute the sensitivity coefficients. The results of the parameter estimation problem are written as $\mathrm{P^{\,\mathrm{est}}}$.\\*
The optimization of cost function is performed through the OPTRAN package~\cite{Dulikravich_1999,Dulikravich_2013}. It includes a hybrid optimizer, combining gradient and global optimization methods to achieve the global extremum of the function. It had the following six constituent optimization modules:
Davidon--Fletcher--Powell(DFP) gradient--based algorithm~\cite{Davidon_1959,Fletcher_1963}, Genetic Algorithm (GA)~\cite{Goldberg_1989}, Nelder--Mead (NM) simplex algorithm~\cite{Nelder_1965}, Differential
Evolution (DE) algorithm~\cite{Storn_1996}, Sequential Quadratic Programming (SQP)~\cite{SQP_1999} and quasi-Newton algorithm of Pshenichny--Danilin (LM)~\cite{LM_1969}. Thus, this hybrid optimizer had three gradient--based and three non--gradient--based constituent optimization algorithms that are automatically switching back--and--forth.


The optimization problem is completed when one of several stopping criterion is met: (1) the maximum number of iterations or objective function
evaluations are exceeded, or (2) the best design in the population was equivalent to a target design, or (3) the optimization program tried all four algorithms but failed to produce a non-negligible decrease in the objective function.

\subsection{Metrics of efficiency and reliability of the model}
\label{sec:sec_ref}
The reliability of the model is assessed by comparing the numerical results with experimental observations. The residual for temperature is computed according to:
\begin{align}
\label{eq:eps2_exp}
\varepsilon\,(\,\chi^{\,\star}_{\,i}\,) \, \eqdef \, 
\Bigl|\,{u^{\,\mathrm{num}}\,\bigl(\,x^{\,\star}\egal \chi^{\,\star}_{\,i}\,\bigr) \moins u_{\,i}^{\,\mathrm{obs}}\,\bigl(\,x^{\,\star}\egal \chi^{\,\star}_{\,i}\,\bigr) }\,\Bigr|
\end{align}
where $\chi^{\,\star}_{\,i}$ is the sensor location, the super script "num" defined the output field computed with the model and "obs" stands for the experimental observation of the field.

Meanwhile, the efficiency of a numerical model can be measured by its computational (CPU) run time required to compute the solution. It is measured using the \texttt{Matlab\texttrademark} environment with a computer equipped with \texttt{Intel} i$7$ CPU and $16$ GB of RAM. 

\section{Case study}
\label{sec:case_study}
\subsection{Presentation}
\begin{figure}[h!]
\begin{center}
\subfigure[\label{fig:wall_real}]{\includegraphics[width=0.45\textwidth]{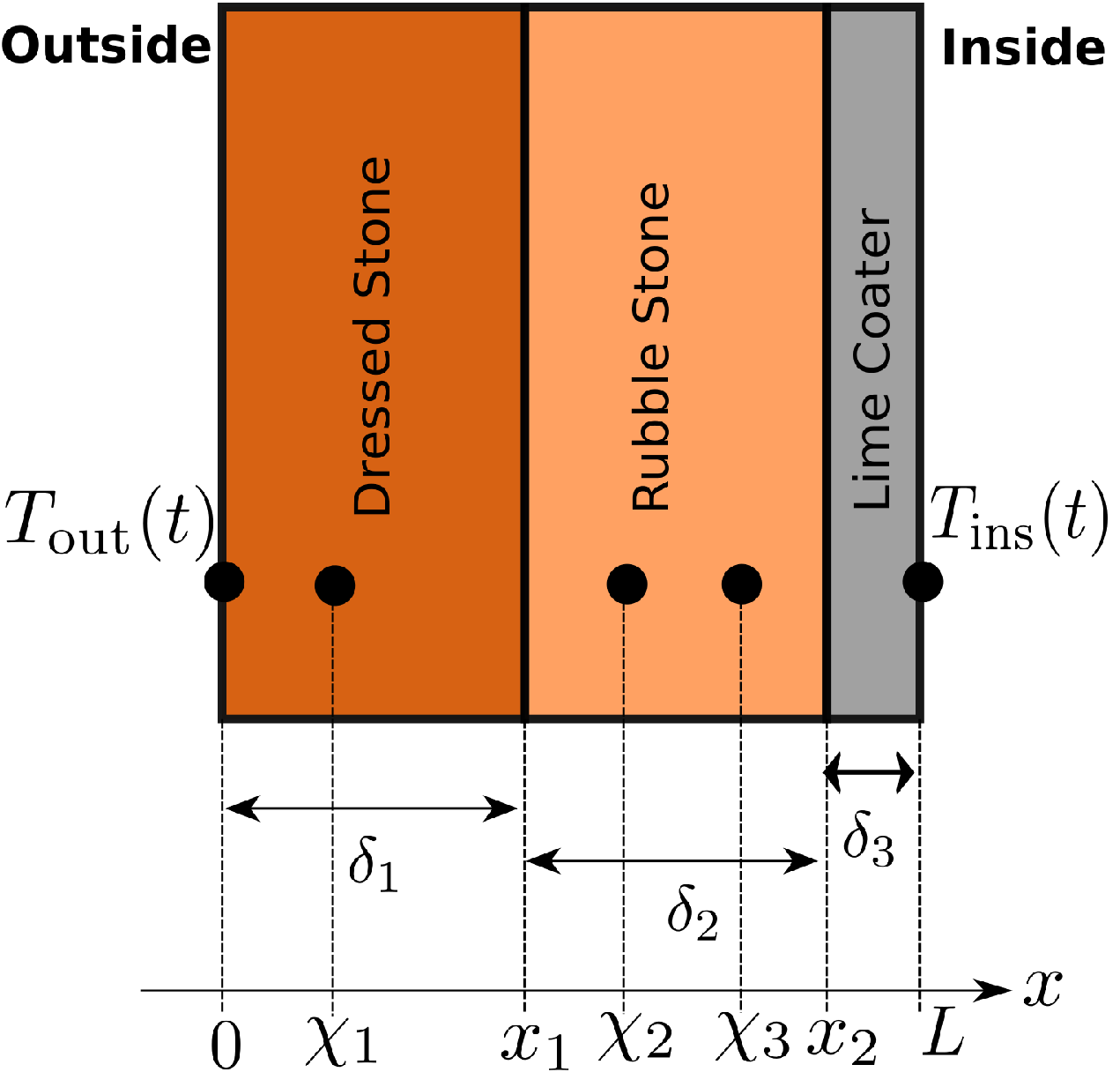}} \hspace{0.25cm}
\subfigure[\label{fig:T_bound}]{\includegraphics[width=0.45\textwidth]{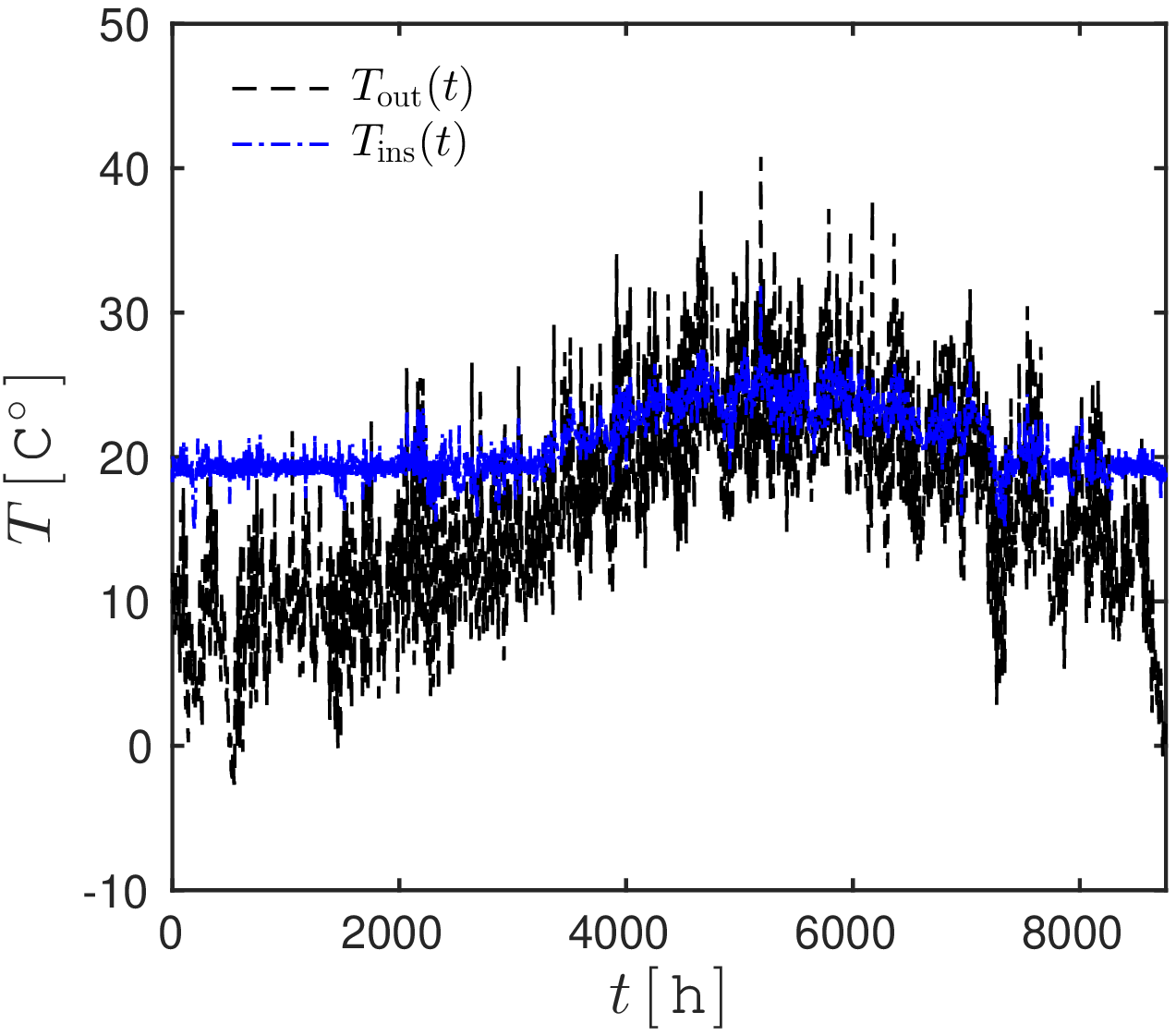}}
\caption{Illustration of the real case study \emph{(a)} with the boundary conditions \emph{(b).}} 
\end{center}
\end{figure}

The issue is to estimate the thermophysical properties of the wall of a  historical building. The house, built in the XIX century, is located in Bayonne, France. The West oriented wall of the living room was considered for the study. The wall is composed of three materials: lime coater, rubble stone and dressed stone. The wall was monitored by sensors which were placed on both sides of the wall surface and three were installed within the wall. The set--up is illustrated in Figure~\ref{fig:wall_real} where $\{\chi_{\,1}\,,\, \chi_{\,2}\,,\, \chi_{\,3}\}$ are the locations of the three sensors. The exact position is reported in Table~\ref{tab:x_observ}.  $T_{\,\mathrm{out}}\,(\,t\,)$ and $T_{\,\mathrm{ins}}\,(\,t\,)$ are outdoor and indoor temperatures. Their time variation is shown in Figure~\ref{fig:T_bound}. 
The thermal properties of the wall are given in Table~\ref{tab:mat_properties},  obtained from the French standards~\cite{fr_std}.  The data acquisition took almost one year starting from the middle of December with a time step of 1 h. Complementary information on the experimental design can be found in~\cite{BERGER_2016}. 

\renewcommand{\arraystretch}{2}

\begin{table}[]
\centering
\caption{Sensors and layers positions within the wall.}
\begin{tabular}{|c|c|c|c|c|c|c|}
\hline
\multirow{2}{*}{Sensors} & $\chi_{\,1}$ $[\,\mathsf{m}\,]$ & $\chi_{\,2}$ $[\,\mathsf{m}\,]$ & $\chi_{\,3}$ $[\,\mathsf{m}\,]$ & \multirow{2}{*}{\shortstack{Layers \\ Interface}} & $x_{\,1}$  $[\,\mathsf{m}\,]$ & $x_{\,2}$  $[\,\mathsf{m}\,]$  \\ \cline{2-4} \cline{6-7} 
                  & 0.05 & 0.23  &0.42  &                   & 0.2 &0.48   \\ \hline
\end{tabular}
\label{tab:x_observ}
\end{table}

\begin{table}[]
\centering
\caption{Thermal properties of each layer.}
\begin{tabular}{|c|c|c|c|c|}
\hline 
Material of layer  & Index   & \shortstack{Thermal conductivity \\ $k_{\,i}^{\,\circ}$ $[\,\mathsf{W\cdot K \cdot m^{-1}}\,]$ }& \shortstack{Heat capacity \\ $c_{\,i}^{\,\circ}$ $[\,\mathsf{J\cdot K^{-1} \cdot m^{-3}}\,]$} & Thickness $\delta_{\,i}$ $[\,\mathsf{m}\,]$ \\ \hline
Dressed Stone & $i \egal 1$  &       $1.75$              &    $1.6 \,\cdot\,10^{\,6} $            &    $0.2$       \\ \hline
Rubble Stone & $i \egal 2$ &     $2.3$                 &       $2.8 \,\cdot\,10^{\,6} $         &     $0.28$      \\ \hline
 Lime Coater & $i \egal 3$ &        $0.8$              &       $2.2 \,\cdot\,10^{\,6} $        &       $0.02$    \\ \hline
\end{tabular}
\label{tab:mat_properties}
\end{table}

\renewcommand{\arraystretch}{1}

\subsection{Experimental observations}
\label{sec:uncert}
The total uncertainty on the observations are evaluated through the propagation of the uncertainties. For the temperature, the total uncertainty is computed according to:
\begin{align}
\sigma \egal \sqrt{\sigma_{\,T}^{\,2} \plus \sigma_{\,x}^{\,2}}
\end{align}
where $\sigma_{\,T} \egal 0.5^{\,\circ}\,C$  is the measurement sensor uncertainty, $\sigma_{\,x}$ is the uncertainty due to the sensor location. The latter is given by following formula:
\begin{align}
\sigma_{\,x} \egal \pd{T}{x}\,\Big|_{x \egal \chi_{\,i}}\,\cdot\,\delta_x\,,
\end{align}
where $\displaystyle{\pd{T}{x}}$ is calculated at the location of the sensors using the numerical model and the \emph{a priori} values of the parameters. The term $\delta_x$ varies according the location, it is $1\,\mathsf{cm}$ when $x\,\in\,\{\,\chi_{\,1}\,,\chi_{\,2}\,,\chi_{\,3}\,\}$, and $\delta_x 
\egal 1.5\,\mathsf{cm}$ if $x\,\in\,\{\,0\,,L\,\}$.

{Several figures present the calculated total uncertainty. Figures~\ref{fig:estim_res2},~\ref{fig:estim_res3},~\ref{fig:estim_res} display the uncertainty in the gray shadow}.

\section{Results of the Parameter Estimation Problem}

\subsection{Structural Identifiability of the Parameters}
\label{sec:struct_id}
The purpose of this section is to demonstrate the theoretical identifiability of the unknown parameters. Moreover, the number of unknown parameters is determined, since it depends on the number of the points of observations and the spatial distribution of thermal conductivity. The evaluation of each case is discussed below.

\subsubsection*{The piecewise function parameterization}
The first demonstration is carried out when thermal conductivity is presented as a piecewise function:
\begin{align}
k^{\,\star}\,(\,x^{\,\star}\,) \egal \begin{cases}
k^{\,\star}_{\,1}\,,\quad 0\ \leqslant \ x^{\,\star} \ < x^{\,\star}_{\,1}\,, \\
k^{\,\star}_{\,2}\,, \quad x^{\,\star}_{\,1} \ \leqslant \ x^{\,\star} \ < \ x^{\,\star}_{\,2} \,,\\
k^{\,\star}_{\,3}\,,\quad x^{\,\star} \ \geqslant \ x^{\,\star}_{\,2} \quad \,.
\end{cases}
\end{align}

First, the parameter $k^{\,\star}_{\,1}$ is considered. Observations $u\,\bigl(\,x^{\,\star}\egal x_{\,1}^{\,\star}\,,t^{\,\star}\,\bigr)\,$ are obtained with the parameter $k^{\,\star}_{\,1}$. Another set of observations 
$u^{\,'}\,\bigl(\,x^{\,\star}\egal x_{\,1}^{\,\star}\,,t^{\,\star}\,\bigr)\,$ is gathered with $k^{\,\star\,'}_{\,1}$.
At the point of observation $x^{\,\star}\egal x_{\,1}^{\,\star}$ the governing equation~\eqref{eq:dim_heat_eq} is as follows:
\begin{align}\label{eq:ident_k1}
c_{\,1}^{\,\star}\;\pd{u}{t^{\,\star}}  \egal \mathrm{Fo}\;\pd{}{x}\,\Bigl(\,k_{\,1}^{\,\star}\;\pd{u}{x^{\,\star}}\,\Bigr)\,.
\end{align}
One may formulate the following equation for the second set:
\begin{align}\label{eq:ident_k1_2}
c_{\,1}^{\,\star}\;\pd{u^{\,'}}{t^{\,\star}}  \egal \mathrm{Fo}\;\pd{}{x}\,\Bigl(\,k_{\,1}^{\,\star\,'}\;\pd{u^{\,'}}{x^{\,\star}}\,\Bigr)\,.
\end{align}
In the case of our model, if $u\,(\,x^{\,\star}\,,t^{\,\star}\,) \ \equiv \ u^{\,'}\,(\,x^{\,\star}\,,t^{\,\star}\,)$, then  $\displaystyle \pd{u}{t^{\,\star}} \ \equiv \ \pd{u^{\,'}}{t^{\,\star}}$ and $\displaystyle \pd{u}{x^{\,\star}}  \ \equiv \ \pd{u^{\,'}}{x^{\,\star}}$. By subtracting equation~\eqref{eq:ident_k1_2} from equation~\eqref{eq:ident_k1}, one may obtain:
\begin{align}
\Bigl(\,k_{\,1}^{\,\star\,'} \moins k_{\,1}^{\,\star}\,\Bigr) \;\pd{u}{x^{\,\star}} \egal 0
\end{align}
Therefore, it can be concluded that $k^{\,\star}_{\,1} \ \equiv \ k^{\,\star\,'}_{\,1}$. 
Similar demonstrations can be carried out to prove that the parameter $k^{\,\star}_{\,2}$ is SGI by using the observations at $x^{\,\star}\egal x_{\,2}^{\,\star}$ or $x^{\,\star}\egal x_{\,3}^{\,\star}$.
However, based on the argumentation above, the parameter $k^{\,\star}_{\,3}$ is not SGI since there are no observations in the third layer.
Therefore, in this particular case only parameters $k^{\,\star}_{\,1}$ and $k^{\,\star}_{\,2}$ can be identified theoretically.

\subsubsection*{Representation through the spline interpolation} 

The next parameterization assumes that thermal conductivity is given by constant values on the first and third layers while changing linearly on the second wall layer. We denote this representation as a linear parameterization. Therefore, the following expression is used:  
\begin{align}
k^{\,\star}\,(\,x^{\,\star}\,) \egal \begin{cases}
k^{\,\star}_{\,1}\,,\quad 0\ \leqslant \ x^{\,\star} \ < x^{\,\star}_{\,1} \,, \\
k^{\star}_{\,20} \plus {\beta}_{\,21}\,x^{\,\star} \,, \quad x^{\,\star}_{\,1} \ \leqslant \ x^{\,\star} \ < \ x^{\,\star}_{\,2} \,,\\
k^{\,\star}_{\,3}\,,\quad x^{\,\star} \ \geqslant \ x^{\,\star}_{\,2} \quad \,.
\end{cases}
\end{align}

Using the same steps from the previous case, one may conclude that parameters  $k^{\,\star}_{\,1}$ and $\{\,k^{\,\star}_{\,20}\,,{\beta}_{\,21}\,\}$  are SGI; the parameter  $k^{\,\star}_{\,3}$ cannot be identified theoretically. 
Indeed, there is only one linear formulation of $k^{\,\star}\,(\,x^{\,\star}\,)$ for the second layer. The two observations at points $\{\,x^{\,\star}_{\,2}\,,\,x^{\,\star}_{\,3}\,\}$ uniquely identify the coefficients $k^{\star}_{\,20}$ and ${\beta}_{\,21}$. 

\subsubsection*{Representation through the polynomial function} 
Another hypothesis implies that thermal conductivity is presented as a second--order polynomial according to the space variable. Therefore,
\begin{align}
k^{\,\star}\,(\,x^{\,\star}\,)\egal k^{\,\star}_{\,00} \plus {\beta}_{\,10}\,x^{\,\star} \plus {\beta}_{\,20}\,x^{\,\star\,2} \,,\qquad \forall \; x^{\,\star} \; \in [\,0\,,1\,]\,.
\end{align}
The theoretical identifiability of $k^{\,\star}\,(\,x^{\,\star}\,)$ is proven below.
Using this function $k^{\,\star}\,(\,x^{\,\star}\,)$  we observe field $u\,(\,x^{\,\star}\,,t^{\,\star}\,)\,:$
\begin{align}
c^{\,\star}\,(\,x^{\,\star}\,)\,\pd{u}{t^{\,\star}} \egal \pd{}{x^{\,\star}}\,\Bigl(\,k^{\,\star}\,(\,x^{\,\star}\,)\,\pd{u}{x^{\,\star}}\,\Bigr)\,
\end{align}
Let us suppose that there is another polynomial $\widehat{k^{\,\star}}\,(\,x^{\,\star}\,)$, and corresponding variable $\widehat{u}\,(\,x^{\,\star}\,,t^{\,\star}\,)\,:$ 
\begin{align}
c^{\,\star}\,(\,x^{\,\star}\,)\,\pd{\widehat{u}}{t^{\,\star}} \egal \pd{}{x^{\,\star}}\,\Bigl(\,\widehat{k^{\,\star}}\,(\,x^{\,\star}\,)\,\pd{\widehat{u}}{x^{\,\star}}\,\Bigr)\,.
\end{align}
Subtracting one equation from another, and using that ${u}\,(\,x^{\,\star}\,,t^{\,\star}\,)\equiv\widehat{u}\,(\,x^{\,\star}\,,t^{\,\star}\,)$, one may obtain:
\begin{align}
\pd{}{x^{\,\star}}\,\Biggl(\,\Bigl(k^{\,\star}\,(\,x^{\,\star}\,)\moins \widehat{k^{\,\star}}\,(\,x^{\,\star}\,)\,\Bigr)\,\pd{u}{x^{\,\star}}\,\Biggr)\egal & 0\,, \qquad \text{or}\\
\Biggl(\,\pd{k^{\,\star}\,(\,x^{\,\star}\,)}{x^{\,\star}} \moins \pd{\widehat{k^{\,\star}}\,(\,x^{\,\star}\,)}{x^{\,\star}} \,\Biggr)\,\pd{u}{x^{\,\star}} \plus \Bigl(k\,(\,x^{\,\star}\,)\moins \widehat{k^{\,\star}}\,(\,x^{\,\star}\,)\,\Bigr)\,\frac{\partial^{\,2}\,u}{\partial\,x^{\,\star\,2}}\egal & 0\,.
\end{align}
The terms $\displaystyle \Bigl\{\,\pd{u}{x^{\,\star}}\,,\frac{\partial^{\,2}\,u}{\partial\,x^{\,\star\,2}}\,\Bigr\}$ are linearly independent. The polynomial $k\,(\,x\,)\moins \widehat{k^{\,\star}}\,(\,x^{\,\star}\,)$ has at most a second order degree. However, it has 3 zero values, the number of observable fields. Therefore, it can be concluded that the polynomial $k^{\,\star}\,(\,x^{\,\star}\,)\moins \widehat{k^{\,\star}}\,(\,x^{\,\star}\,)$ is a zero polynomial, and $k^{\,\star}\,(\,x^{\,\star}\,)\ \equiv \ \widehat{k^{\,\star}}\,(\,x^{\,\star}\,)\,.$ 
So all three parameters $\{\,k^{\,\star}_{\,00}\,,\,{\beta}_{\,10}\,,{\beta}_{\,20}\,\}$ are SGI.\\

The choice of the thermal conductivity spatial distribution can be questioned. The answer lies in the number of available sensors. In this particular case, at most three parameters can be theoretically identified. In order to use more complex formulations of the thermal conductivity parameterization, the experimental design should provide more observable fields. For example, to increase an order of the polynomial parameterization and use a cubic polynomial, one has to put one more sensor inside the wall. 
With the given theoretical identifiablity, one may search the Optimal Experiment Design to ensure a high quality parameter estimation.

\subsection{Results of the Optimal Experiment Design}
\label{sec:oed}

Aim of this section is to demonstrate how to select an optimal measurement plan, which allows us to estimate the thermal conductivity accurately and efficiently. Furthermore, it can be remarked that the search of the OED requires only the knowledge of wall boundary conditions. So the methodology is reproducible and does not need measurements inside the wall.
  
\begin{figure}[h!]
\begin{center}
\subfigure[\label{fig:F_1day}]{\includegraphics[width=0.45\textwidth, height=7cm]{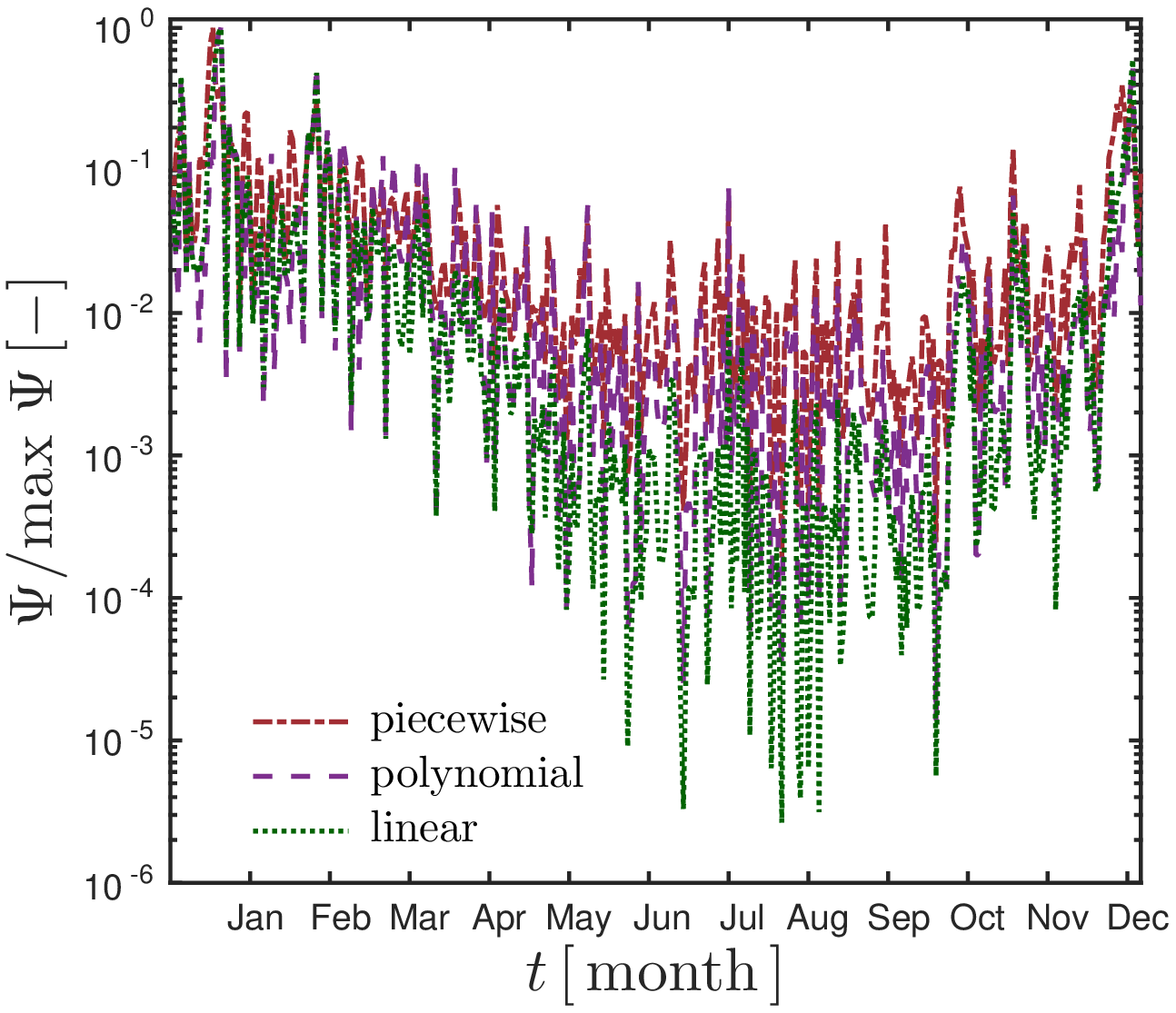}} \hspace{0.2cm}
\subfigure[\label{fig:F_3day}]{\includegraphics[width=0.45\textwidth, height=7cm]{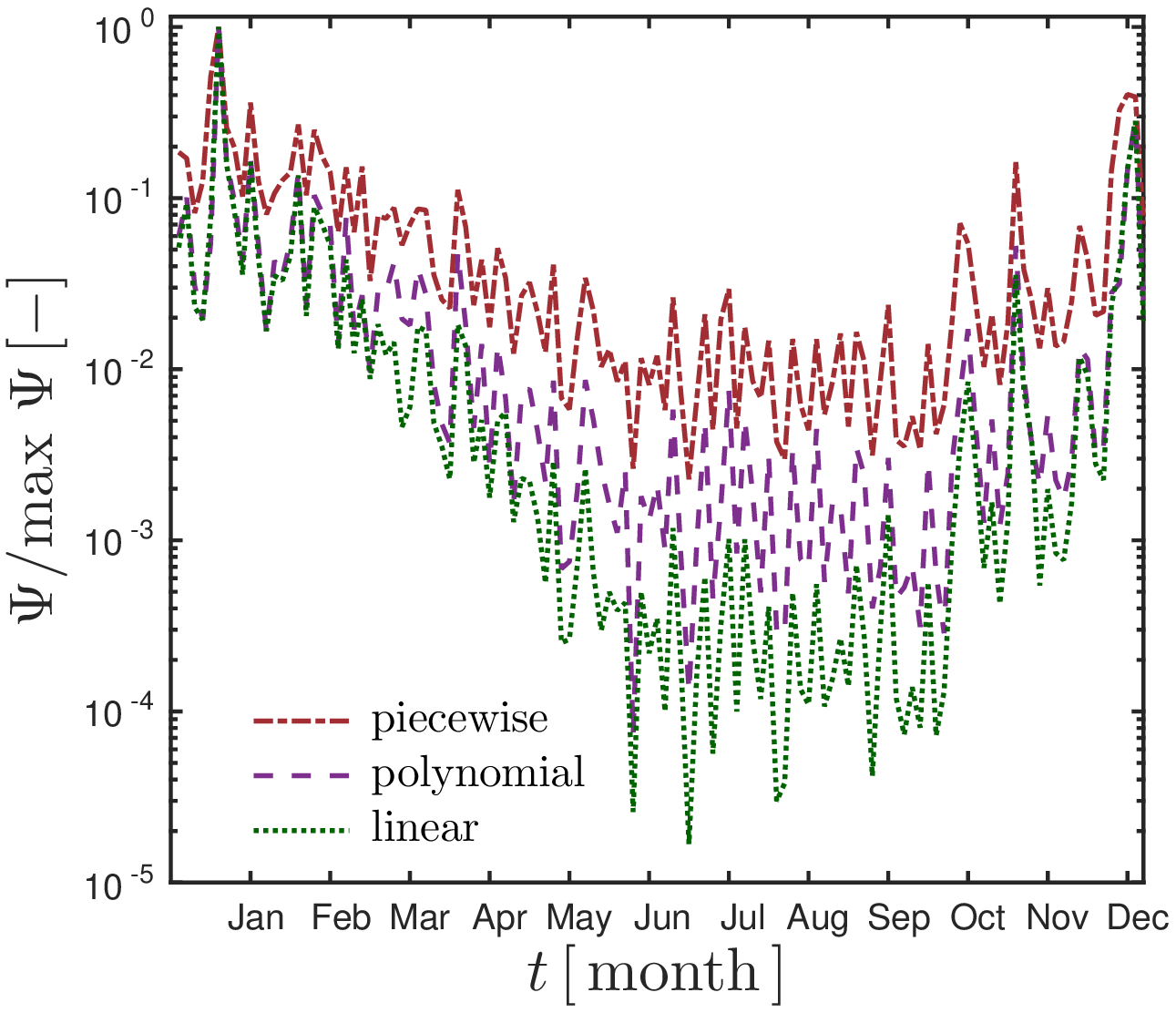}} \quad
\subfigure[\label{fig:F_7day}]{\includegraphics[width=0.45\textwidth, height=7cm]{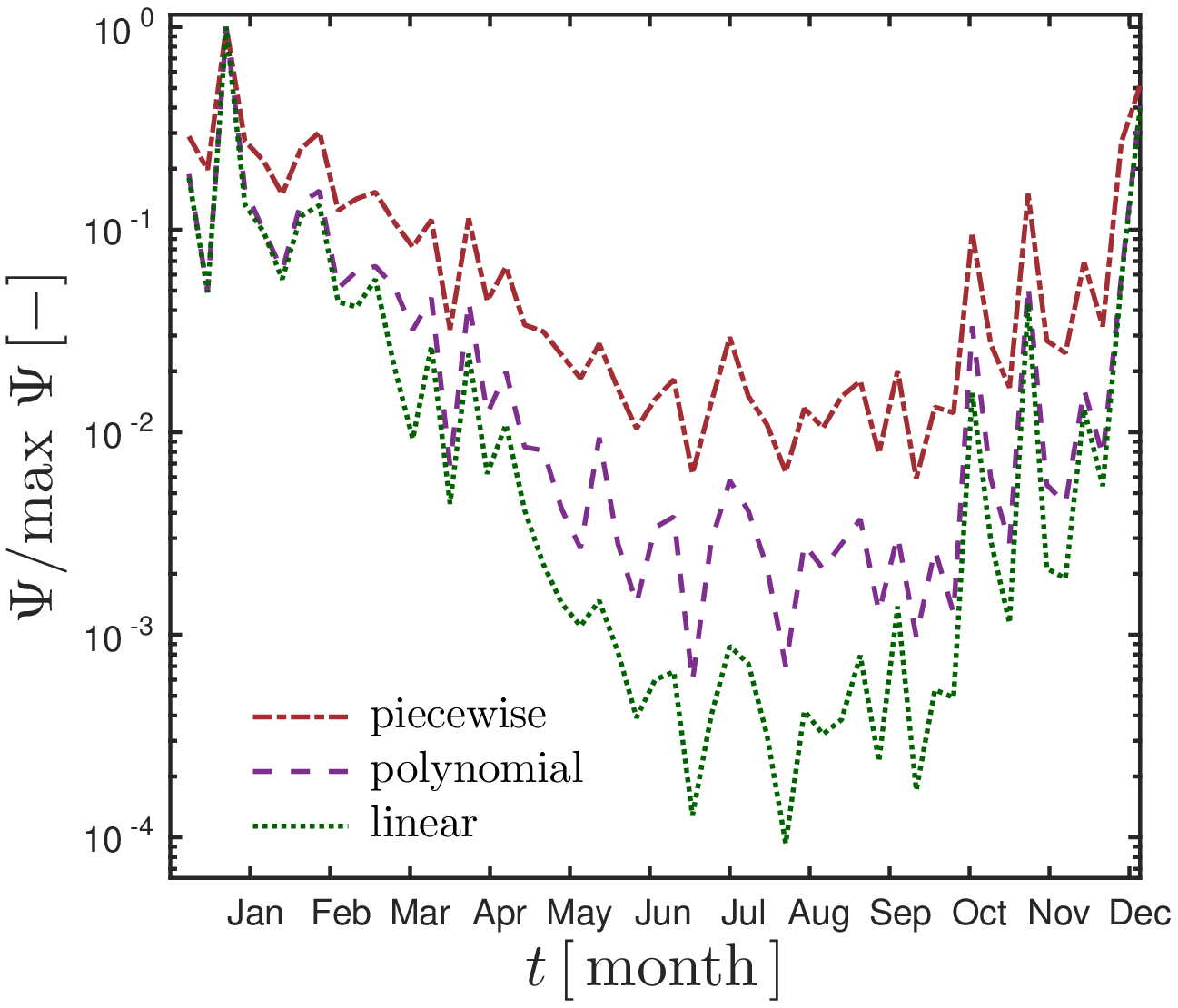}} 
\caption{Variation of the criteria $\Psi$ according to the length of observation period $\delta\,\tau \egal 1 \ \mathsf{days}$ \emph{(a)},  $\delta\,\tau \egal 3 \ \mathsf{days}$ \emph{(b)} and $\delta\,\tau \egal 7 \ \mathsf{days}$ \emph{(c)}.}
\label{fig:OED_detF}
\end{center}
\end{figure}

The three different measurement plans are investigated:
\begin{align}
& t_{\,\mathrm{ini}} \ \in \ [\,0\,,360\,]\quad [\,\mathsf{day(s)}\,]\,,\\
& \delta\,\tau  \egal 1\,, &&  \delta\,\tau \egal 3\,, &&&  \delta\,\tau \egal 7  \,, \nonumber \\ 
& N_{\,\delta\,\tau} \egal 360\,,&& N_{\,\delta\,\tau}  \egal 120\,, &&&  N_{\,\delta\,\tau}  \egal 52\,. \nonumber
\end{align} 
To calculate the modified \textsc{Fisher} matrix using Eq.~\eqref{eq:fisher} the sensitivity coefficients of the particular parameters are required. Thus, for each measurement plan, three thermal conductivity parametrizations are studied. 
The sensitivity equations of the piecewise, linear and polynomial (quadratic) representation  are solved with the~\DF\ numerical scheme. 
For the OED investigation the \emph{a priori} parameter values used are reported in Table~\ref{tab:mat_properties}.
The results are summarized in Table~\ref{tab:oed_res} and Figure~\ref{fig:OED_detF}. Table~\ref{tab:oed_res} presents the starting dates, from which the D--optimum criterion reached its maximum and minimum values depending on the duration of the experiment and the parameterization. Table~\ref{tab:oed_res} shows that to obtain the best accuracy of estimated parameters the observations in January should be chosen. This is valid for each thermal conductivity representation and each duration of the measurement plan. However, the observations in the summer period generally give a lower accuracy of estimated parameters. 

\begin{table}[]
\centering
\caption{The values of initial day $t_{\,\mathrm{ini}}$, starting from which the minimum and maximum of $\Psi$  are achieved.}
\begin{tabular}{|c|c|c|c|c|c|c|}
\hline
\multirow{2}{*}{ \emph{\shortstack{Type of \\parametrization}}} & \multicolumn{2}{c|}{$\delta\,\tau \egal 1 \,[\,\mathsf{day}\,]$} & \multicolumn{2}{c|}{$\delta\,\tau \egal 3 \,[\,\mathsf{days}\,]$} & \multicolumn{2}{c|}{$\delta\,\tau \egal 7 \,[\,\mathsf{days}\,]$} \\ \cline{2-7} 
                  &       $\max\,\Psi$    &   $\min\,\Psi$       &        $\max\,\Psi$    &   $\min\,\Psi$      &        $\max\,\Psi$    &   $\min\,\Psi$        \\ \hline
     \emph{Piecewise}             &     January, 16th       &    August, 18th       & January, 18th           &         July, 14th &   January, 21st        &  August, 5th                    \\ \hline
       \emph{Linear}            &   January, 19th        &    August, 18th             &  January, 18th         &          July, 14th &    January, 21st      &       August, 8th                \\ \hline
        \emph{Polynomial}           &    January, 19th      &   September, 2nd        &   January, 18th        &          June, 23rd &  January, 21st         &    September, 2nd      \\ \hline
\end{tabular}
\label{tab:oed_res}
\end{table}

Figure~\ref{fig:OED_detF} displays values of the relative D--optimum criterion depending on the selection of the measurement plan. 
Figure~\ref{fig:F_3day} demonstrates how the relative criterion composed of 120 values varies during the whole year of the experimental campaign. It can be noted that the high values of the criterion occurred in the cold season, providing a better accuracy for the  parameter estimation problem, while the lower values occurred during the summer. Moreover, the D--optimality criterion of the three thermal conductivity parametrizations can be compared. Figure~\ref{fig:F_7day} displays that the piecewise representation had the highest values of the criterion, showing that piecewise parameters are more sensitive than the others. 

Furthermore, a difference between the highest values of the criterion for three chosen sequences can be discussed. Table~\ref{tab:oed_all} provides an information how the criterion varies depending on the measurement plan using the piecewise representation. It can be seen that $7 \,[\,\mathsf{days}\,]$ of observation provide more information than the others since it provides the highest value of the objective function $\Psi$.  
\begin{table}[]
\centering
\caption{The D-criterion highest values of the piecewise parametrization.}
\begin{tabular}{|c|c|c|c|}
\hline
 & $\delta\,\tau \egal 1 \,[\,\mathsf{day}\,]$  & $\delta\,\tau \egal 3 \,[\,\mathsf{days}\,]$ &  $\delta\,\tau \egal 7 \,[\,\mathsf{days}\,]$ \\ \hline
 $\max\,\Psi$ & $6.7\,\times\,\mathcal{O}\,(\,-13\,)\,$ & $3.9\,\times\,\mathcal{O}\,(\,-11\,)\,$  & $3.2\,\times\,\mathcal{O}\,(\,-10\,)\,$  \\ \hline
\end{tabular}
\label{tab:oed_all}
\end{table}

The decision on the duration of the experiments is a difficult compromise. A week of observations might be chosen since it ensures maximum accuracy, when solving the inverse problem, compare to one-day duration.  However, the cost of the inverse problem increases exponentially with the duration of the experiment. Thus, a duration of $3 \,[\,\mathsf{days}\,]$ for the experiments seems a good compromise as a measurement plan. Moreover, since we are dealing with an occupied building, the experiment should not disturb the occupants. 
The next step is to study the practical identifiability of parameters.

\subsection{Results of Practical Identifiability}
\begin{figure}[h!]
\vspace*{-0.5cm}
\begin{center}
\subfigure[$t_{\,\mathrm{ini}} \egal 18/01$ , $x^{\,\star}\egal \chi^{\,\star}_{\,1}$\label{fig:sc_jan_1}]{\includegraphics[width=0.45\textwidth, height=7.5cm]{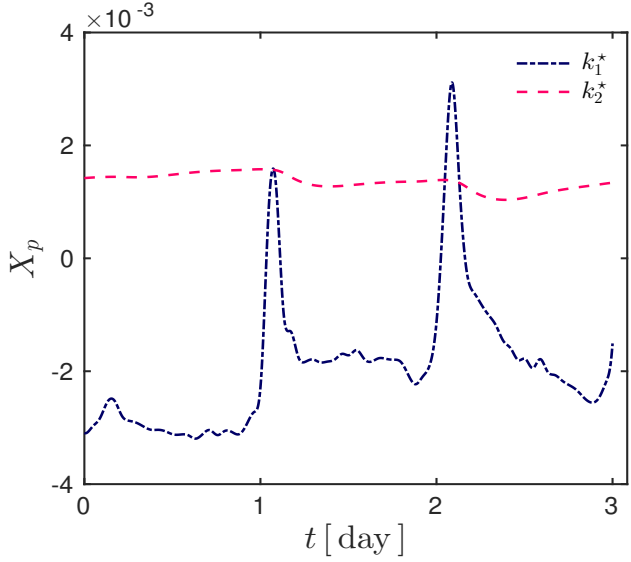}} 
\hspace{0.2cm}
\subfigure[$t_{\,\mathrm{ini}} \egal 14/07$ , $x^{\,\star}\egal \chi^{\,\star}_{\,1}$\label{fig:sc_june_1}]{\includegraphics[width=0.45\textwidth, height=7.5cm]{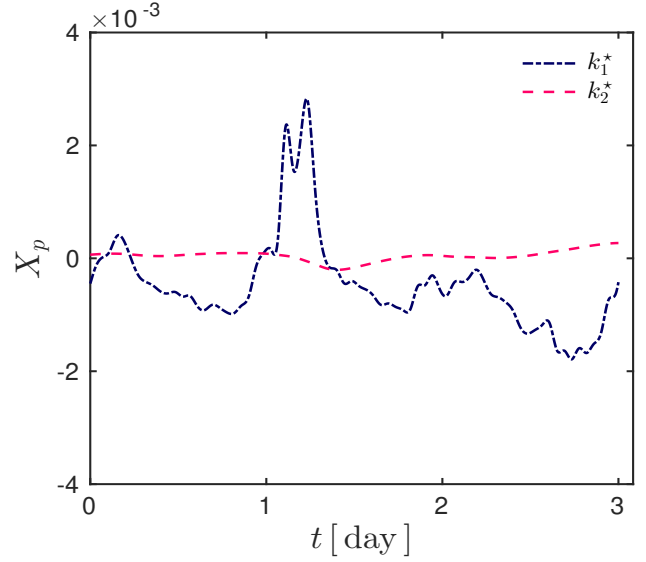}} 
\quad
\subfigure[$t_{\,\mathrm{ini}} \egal 18/01$ , $x^{\,\star}\egal \chi^{\,\star}_{\,2}$\label{fig:sc_jan_2}]{\includegraphics[width=0.45\textwidth, height=7.5cm]{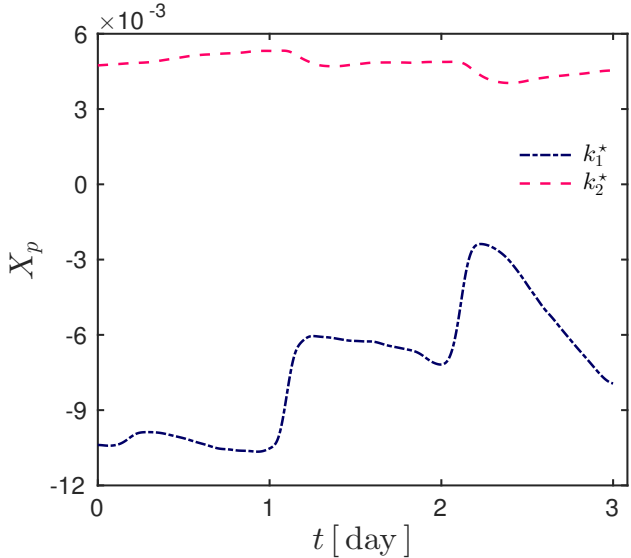}} 
\hspace{0.2cm}
\subfigure[$t_{\,\mathrm{ini}} \egal 14/07$, $x^{\,\star}\egal \chi^{\,\star}_{\,2}$\label{fig:sc_june_2}]{\includegraphics[width=0.45\textwidth, height=7.5cm]{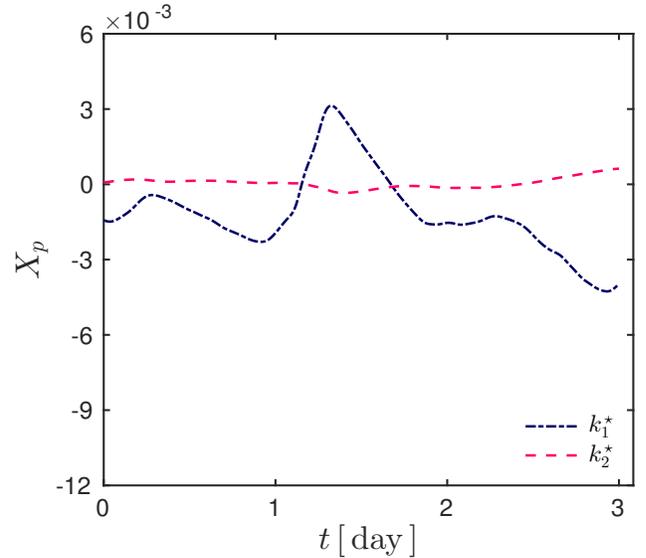}} 
\caption{Variation of the sensitivity coefficients at the sensors location during 3 days in January and July. }
\label{fig:SC_Jan_June}
\end{center}
\end{figure}

The issue now is to demonstrate the practical identifiability of the parameters for the selected three days of observations, $t_{\,\mathrm{ini}} \egal 18/01\,,\ \delta\,\tau \egal 3$ . For the sake of compactness, the sensitivity coefficients are presented only for the piecewise thermal conductivity. Furthermore, as discussed in Section~\ref{sec:oed}, the piecewise parameters are more sensitive.
Section~\ref{sec:struct_id} shows that the two parameters $k^{\,\star}_{\,1}$ and $k^{\,\star}_{\,2}$ are theoretically identifiable for piecewise representation. Therefore, only two sensitivity functions were computed. The calculation was performed for the whole year. Next, considering the different periods of the year, reported in Table~\ref{tab:oed_res}, the variation over the time of the variables $X_{\,k^{\,\star}_{\,1}}$ and $X_{\,k^{\,\star}_{\,2}}$ is shown in Figure~\ref{fig:SC_Jan_June}. Figures~\ref{fig:sc_jan_1} and ~\ref{fig:sc_jan_2} display the variation of the sensitivity coefficients during three days in January for the first two sensors. In addition, Figures~\ref{fig:sc_june_1} and ~\ref{fig:sc_june_2} present the same information during a three day period in July corresponding to the period of the experimental design with the lowest values of D--criterion. Several conclusions can be drawn.

First, it can be remarked that, in general, the amplitudes of the sensitivity coefficients during the winter period were higher than in the summer. As a result, the model is more sensitive to the parameters during the chosen period in January. This is consistent with the fact that the $\mathrm{D}$--optimum criterion achieved its highest value in January. 
Secondly, the sensitivity coefficients are linearly independent within the first and second layers. Therefore, the thermal conductivity of the first and second layers are practically identifiable. As shown in Figures~\ref{fig:sc_jan_1} and~\ref{fig:sc_jan_2}, the sensitivity of the parameter $k^{\,\star}_{\,1}$ is higher for the sensor located in the first layer. Additionally, Figures~\ref{fig:sc_jan_1} and~\ref{fig:sc_jan_2} show the sensitivity function of parameter $k^{\,\star}_{\,2}$ increases for the sensor position in the second layer.  Moreover, the total sensitivity was calculated during the best period of observations between the two parameters in all sensor locations. The following expression can be used:
\begin{align}
\mathrm{F} \,\Bigl(\,k^{\,\star}_{\,1}\,,\,k^{\,\star}_{\,2}\,\Bigr) & \egal \displaystyle \sum_{q\egal 1}^{3}\nicefrac{1}{\sigma^{\,2}}\, \int_{\,\delta\tau}{ X_{\,k^{\,\star}_{\,1}}\,\big(\,\chi_{\,q}\,,\,t\,\big) \, X_{\,k^{\,\star}_{\,2}}\,\big(\,\chi_{\,q}\,,\,t\,\big) \; \mathrm{d}t}\,.
\end{align}
For this particular case this value is $-0.0029$, which proves the sensitivity coefficients are not correlated.
It can be concluded that the unknown parameters $k^{\,\star}_{\,1}$ and $k^{\,\star}_{\,2}$ are identifiable from the practical aspect, and should be estimated during the winter season. 
\begin{align}
\label{mat:Fisher}
\mathrm{F}_{\,\mathrm{lin}} \egal \bordermatrix{& k^{\,\star}_{\,1} & k^{\,\star}_{\,20} & \beta_{\,21} \cr k^{\,\star}_{\,1} & 1 & -0.0028 & -0.0022 \cr k^{\,\star}_{\,20} & - & 1 & 0.0014 \cr \beta_{\,21} & - & - & 1}\,, \quad
\mathrm{F}_{\,\mathrm{quad}} \egal \bordermatrix{& k^{\,\star}_{\,1} & \beta_{\,10} & \beta_{\,20} \cr k^{\,\star}_{\,1} & 1 & 0.0020 & 0.0046 \cr \beta_{\,10} & - & 1 & 0.0020 \cr \beta_{\,20} & - & - & 1}\,.
\end{align}

Similarly, the total sensitivity can be computed for linear and polynomial thermal conductivity representation. The results are reported in matrices (~\ref{mat:Fisher}). The matrices values do not equal to the zero value, so, the parameters can be estimated practically.

\subsection{Results of the Parameter Estimation Problem}

The parameter estimation problem is solved using the optimization procedure described in Section~\ref{sec:cost_min} and the direct numerical model detailed in Section~\ref{sec:num_mod}. The space and time discretization are $\Delta x^{\,\star}\egal 10^{-2}$ and $\Delta t^{\,\star}\egal 10^{-2}$ corresponding, from a physical point of view, to $\Delta x \egal 5\,\cdot\,10^{-3}\, \mathsf{m}$ and $\Delta t \egal 3.6\,\mathsf{s}$. The direct problem is solved for three days in January, $t_{\,\mathrm{ini}} \egal 18/01\,,\ \delta\,\tau \egal 3$  as justified in Section~\ref{sec:oed}. The three separate cases of thermal conductivity parameterization are analyzed. 
\begin{table}[]
\centering
\caption{The \emph{a priori} and estimated parameter values of the thermal conductivity representations.}
\begin{tabular}{|c|c|c|}
\hline
Parameter & \emph{A priori} value & Estimated value    \\ \hline
\multicolumn{3}{|c|}{\textbf{Piecewise}}  \\ \hline
$k^{\star\,\circ}_{\,1}$ & $1.0$ & $0.75$     \\ \hline
$k^{\star\,\circ}_{\,2}$ &  $1.3$ & $1.01$     \\ \hline
\multicolumn{3}{|c|}{\textbf{Linear}} \\ \hline
$k^{\star}_{\,1}$ & $1.0$ & $0.8336$  \\ \hline
$k^{\star}_{\,20}$ &  $0.6195$ & $-0.714$ \\ \hline
$\beta_{\,21}$ &  $0.8271$ & $2.639$  \\ \hline  
\multicolumn{3}{|c|}{\textbf{Quadratic}} \\ \hline
$k^{\star}_{\,00}$ & $0.89$ & $1.3952$   \\ \hline
$\beta_{\,10}$ &  $2.397$ & $-3.8249$ \\ \hline
$\beta_{\,20}$ &  $-2.655$ & $4.4296$  \\ \hline
\end{tabular}
\label{tab:k_est_all}
\end{table}

\begin{figure}[h!]
\begin{center}
\subfigure[$t_{\,\mathrm{ini}} \egal 18/01$ , $x \egal \chi_{\,1}$\label{fig:estim3_1}]{\includegraphics[width=0.45\textwidth, height=7.5cm]{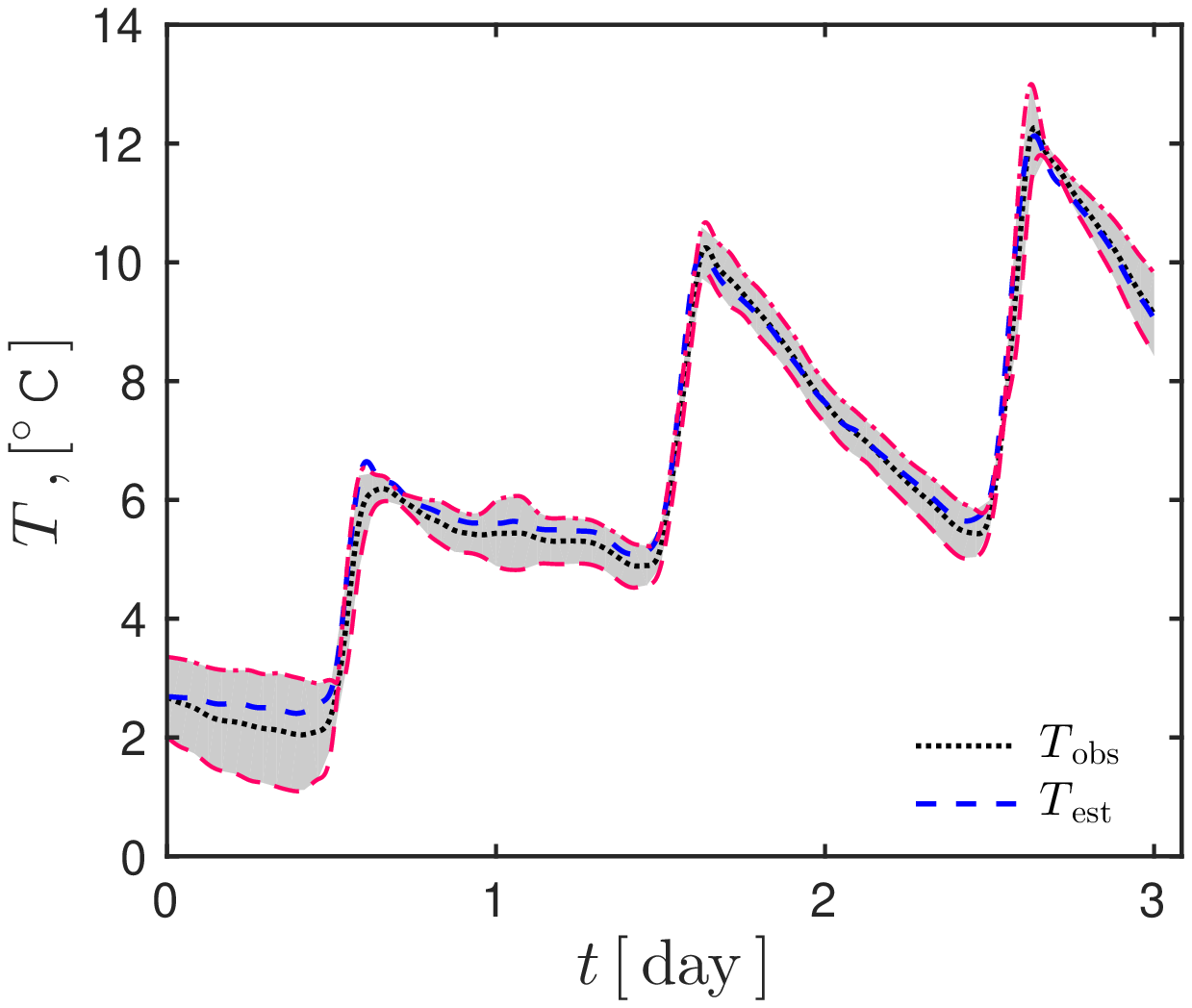}} \quad
\subfigure[$t_{\,\mathrm{ini}} \egal 18/01$ , $x \egal \chi_{\,2}$\label{fig:estim3_2}]{\includegraphics[width=0.45\textwidth, height=7.5cm]{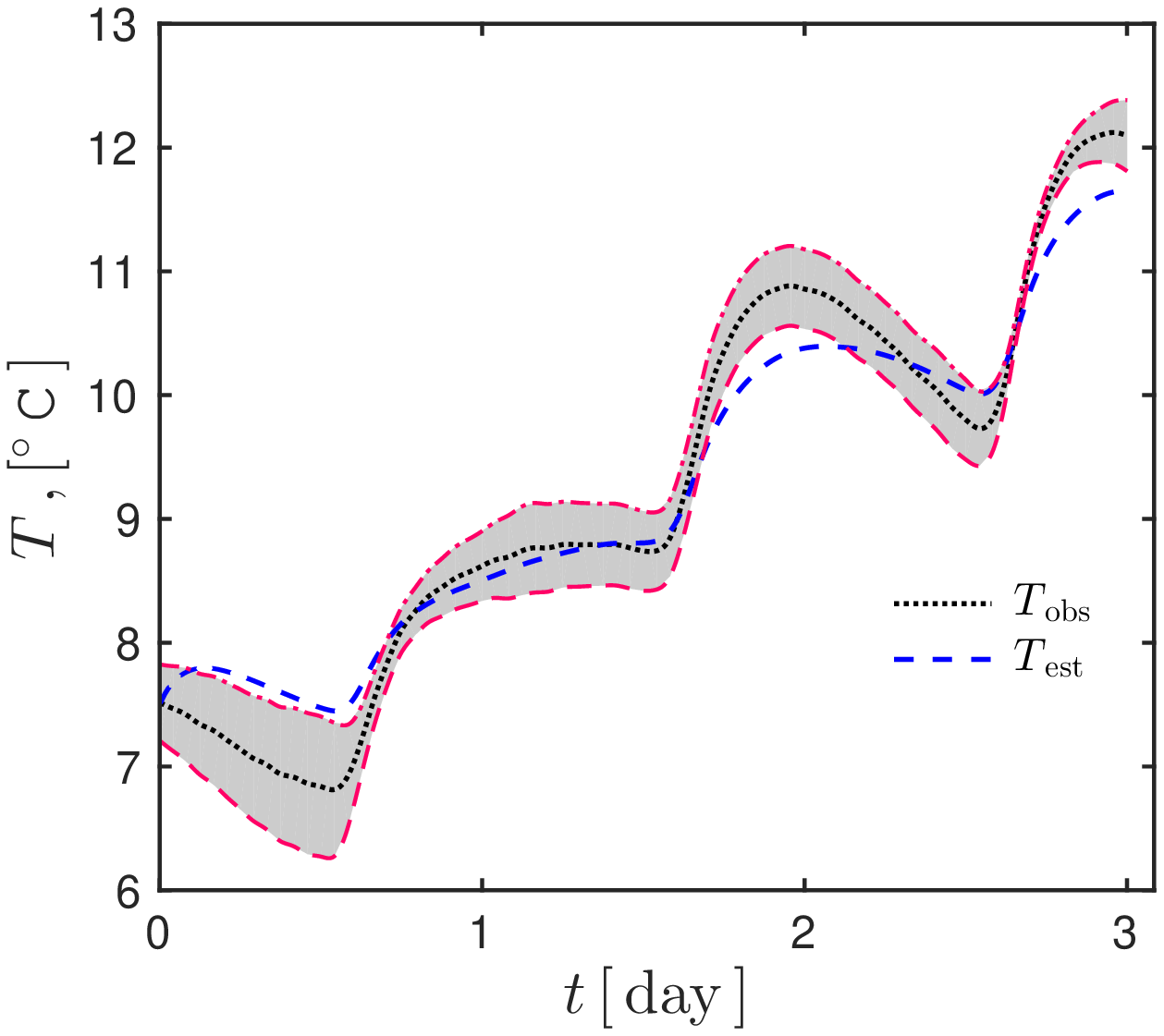}} \quad
\subfigure[$t_{\,\mathrm{ini}} \egal 18/01$ , $x \egal \chi_{\,3}$\label{fig:estim3_3}]{\includegraphics[width=0.45\textwidth, height=7.5cm]{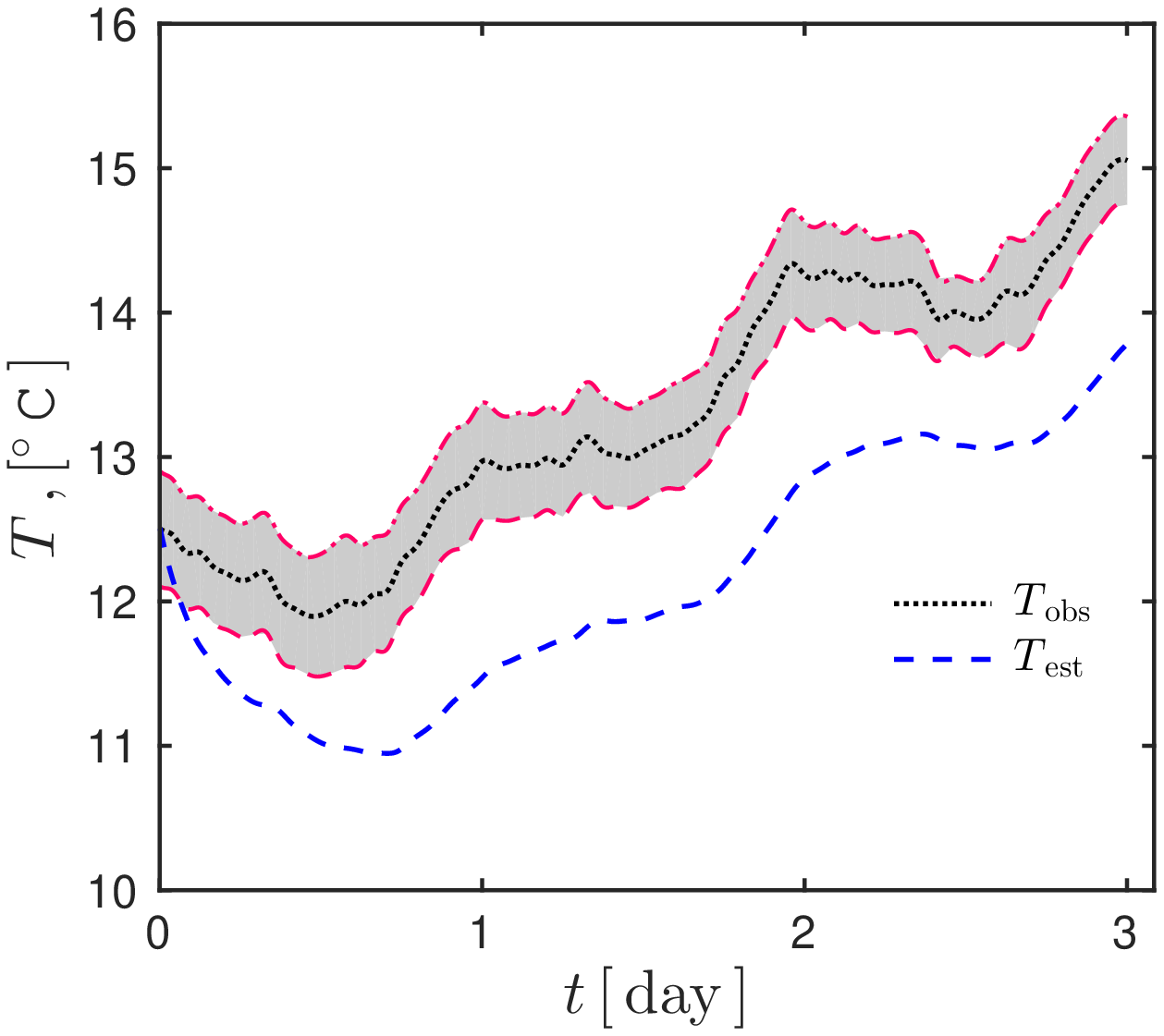}} \quad
\subfigure[\label{fig:k_step}]{\includegraphics[width=0.45\textwidth, height=7.5cm]{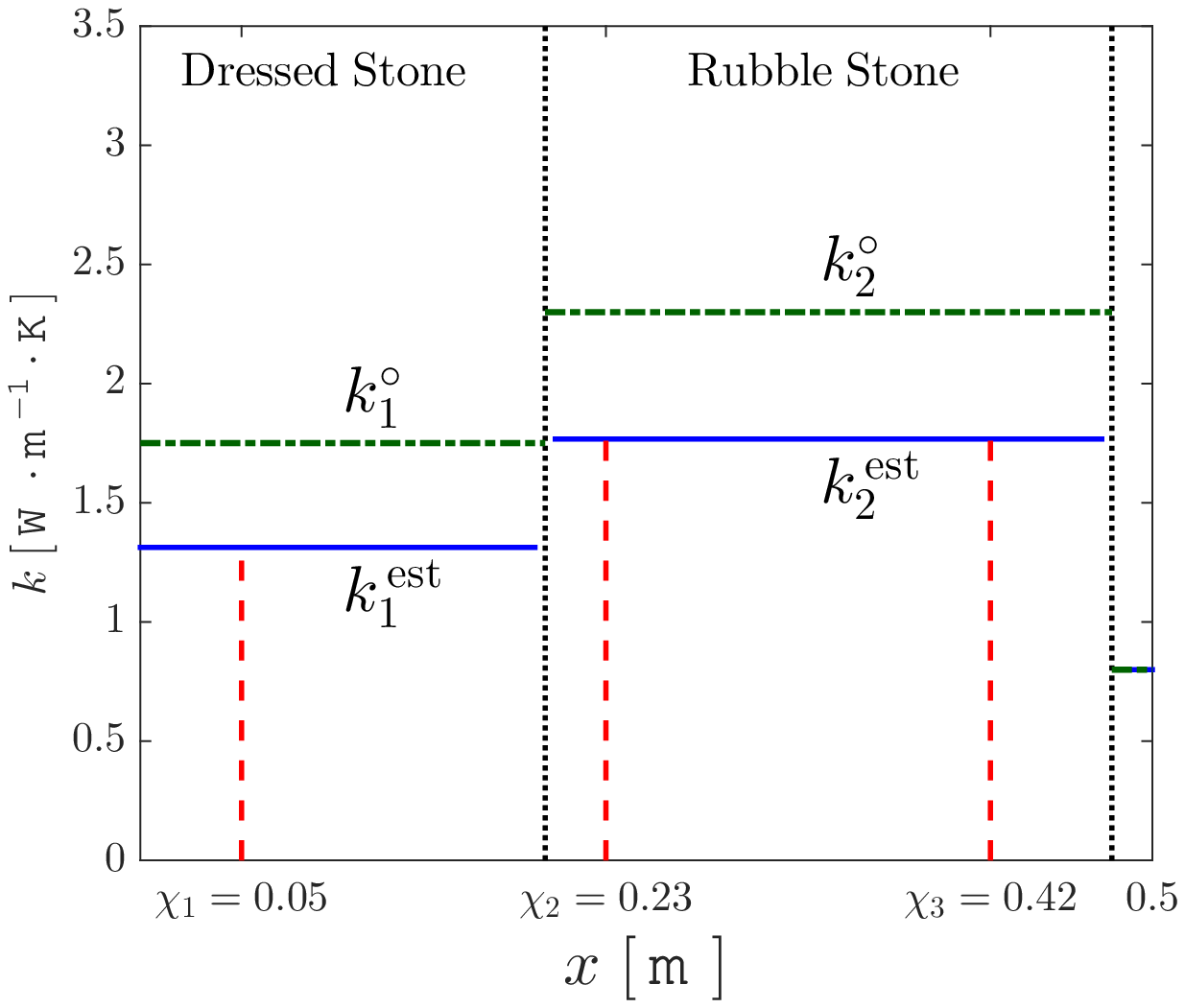}}
\caption{Variation of the computed temperature and the experimental observations for \emph{(a)} first ,\emph{(b)} second and \emph{(c)} third sensor locations; and \emph{(d)} the estimated thermal conductivity representation.}
\label{fig:estim_res3}
\end{center}
\end{figure}

First, the piecewise representation of the thermal conductivity was studied. As a result of the theoretical identifiability, only 2 parameters $k^{\star\,\circ}_{\,1}$ and $k^{\star\,\circ}_{\,2}$ were identified. The initial guess and the estimated results are reported in Table~\ref{tab:k_est_all}.
Then the direct problem was solved with the estimated parameter values and compared to the observations.
The calculated temperature values and the respective measurements with uncertainty boundaries in three sensors locations are presented in Figure~\ref{fig:estim_res3}. The estimation was in good agreement for the first two sensors. However, in the third sensor location, the computed temperature values were out of the observation uncertainty bound. This discrepancy may arise due to the modeling of the second layer material properties. As rubble stone is not a homogeneous material, and it is irregular in shape and structure.  
Figure~\ref{fig:k_step} displays the estimated variation of thermal conductivity according to the space coordinate and the standard values of the thermal conductivity.

\begin{figure}[h!]
\begin{center}
\subfigure[$t_{\,\mathrm{ini}} \egal 18/01$ , $x \egal \chi_{\,1}$\label{fig:estim2_1}]{\includegraphics[width=0.45\textwidth, height=7.5cm]{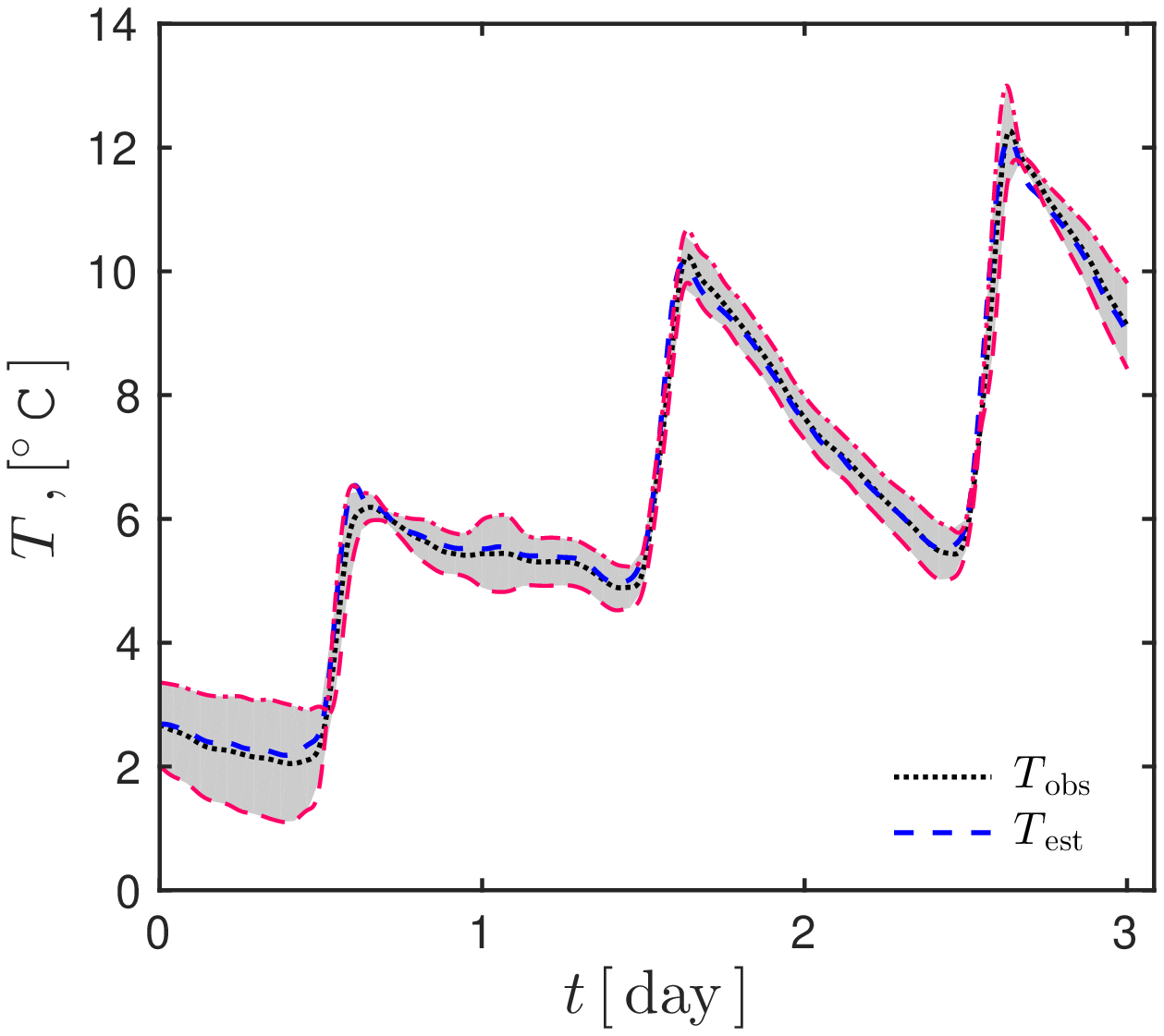}} \quad
\subfigure[$t_{\,\mathrm{ini}} \egal 18/01$ , $x \egal \chi_{\,2}$\label{fig:estim2_2}]{\includegraphics[width=0.45\textwidth, height=7.5cm]{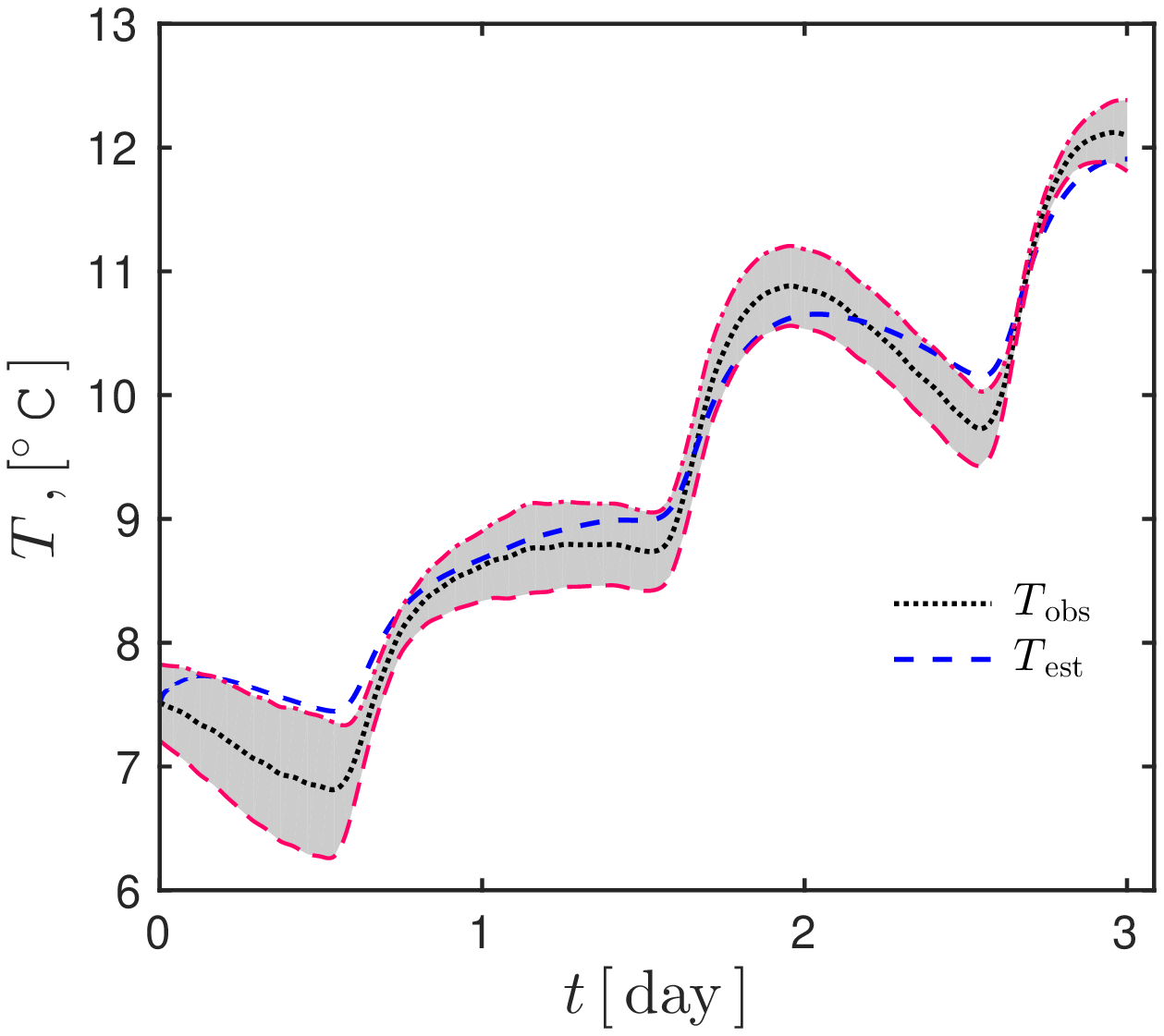}} \quad
\subfigure[$t_{\,\mathrm{ini}} \egal 18/01$ , $x \egal \chi_{\,3}$\label{fig:estim2_3}]{\includegraphics[width=0.45\textwidth, height=7.5cm]{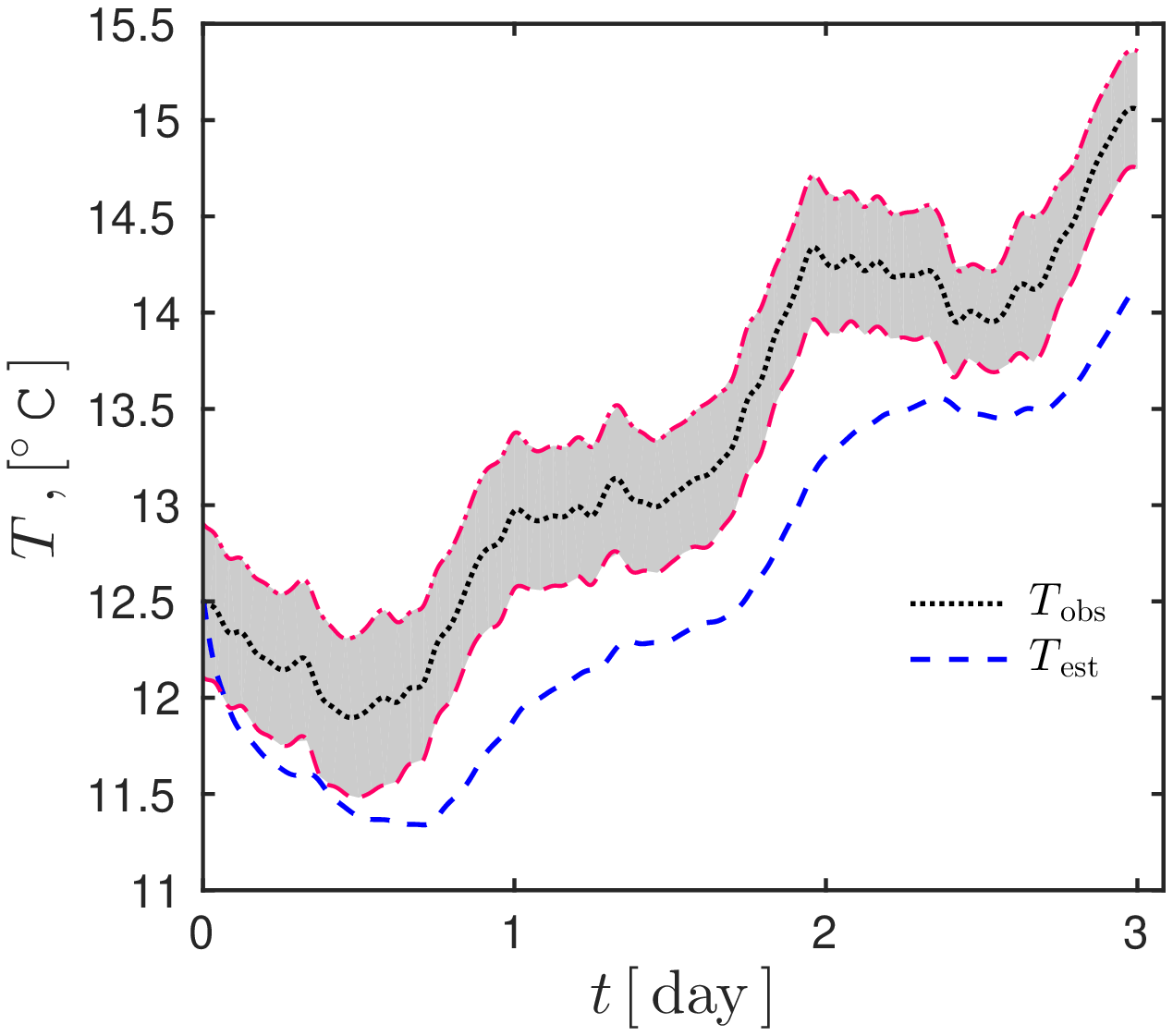}} \quad
\subfigure[\label{fig:k_lin}]{\includegraphics[width=0.45\textwidth, height=7.5cm]{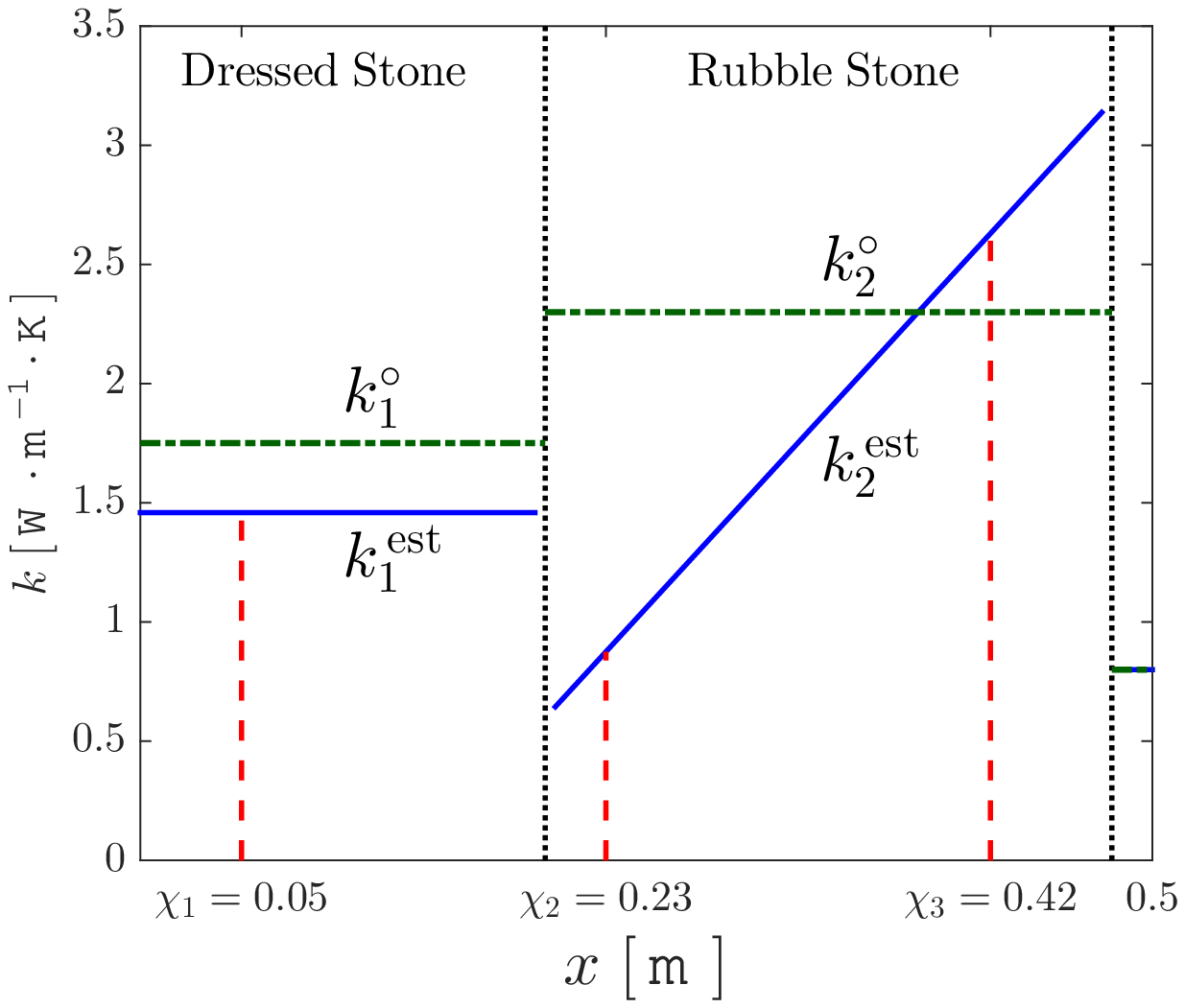}}
\caption{ Variation of the computed temperature and the experimental observations for \emph{(a)} first ,\emph{(b)} second and \emph{(c)} third sensor locations; and \emph{(d)} the estimated thermal conductivity representation.}
\label{fig:estim_res2}
\end{center}
\end{figure}

In the second case, the parameters of the linear representation were identified. The parameter scope includes the thermal conductivity of the first layer $k^{\star}_{\,1}$ and the coefficients $\{\,k^{\star}_{\,20}\,,\beta_{\,21}\,\}$ of the linear function for the second layer. The initial guess and the estimated results are reported in Table~\ref{tab:k_est_all}.
Figure~\ref{fig:estim_res2} illustrates the difference between observations and the computed temperature values in different sensor locations. Similar to the previous case, the calculated temperature values are closer to the measurements in the first two sensors. However, a comparison of Figure~\ref{fig:estim2_3} and Figure~\ref{fig:estim3_3} shows that the gap between the experimental and the calculated data was smaller for the linear representation. Although, the computed values were not in the measurement uncertainty bounds.   
Figure~\ref{fig:k_lin} presents the variation of the estimated thermal conductivity according to the location in the wall and a comparison with the standard values of the thermal conductivity. It can be seen that the thermal conductivity of the second layer was higher closer to the end of the wall. 

\begin{figure}[h!]
\begin{center}
\subfigure[$t_{\,\mathrm{ini}} \egal 18/01$ , $x \egal \chi_{\,1}$\label{fig:estim_1}]{\includegraphics[width=0.45\textwidth, height=7.5cm]{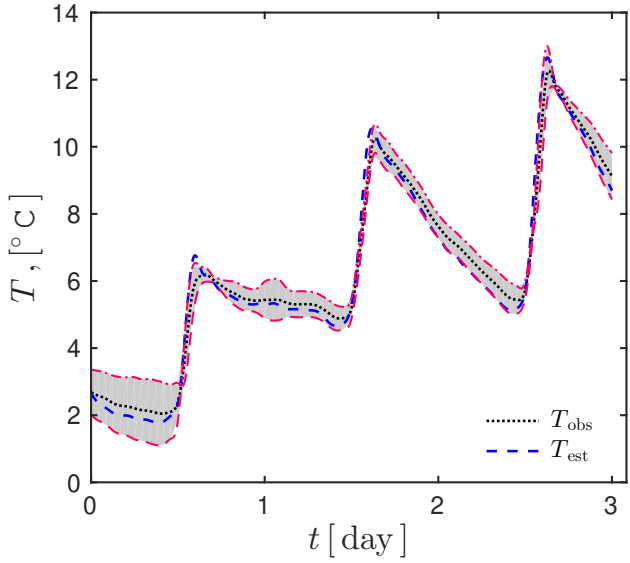}} \quad
\subfigure[$t_{\,\mathrm{ini}} \egal 18/01$ , $x \egal \chi_{\,2}$\label{fig:estim_2}]{\includegraphics[width=0.45\textwidth, height=7.5cm]{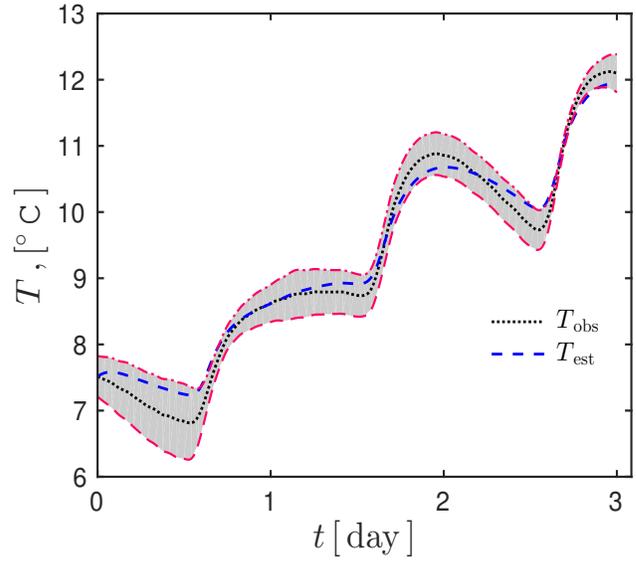}} \quad
\subfigure[$t_{\,\mathrm{ini}} \egal 18/01$ , $x \egal \chi_{\,3}$\label{fig:estim_3}]{\includegraphics[width=0.45\textwidth, height=7.5cm]{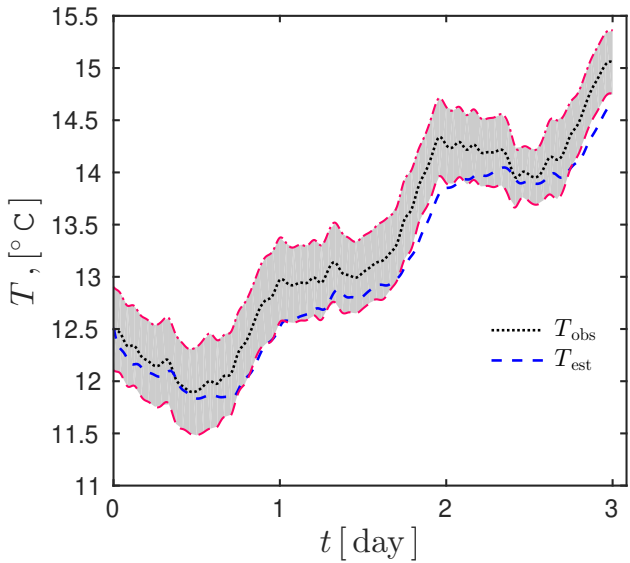}} \quad
\subfigure[\label{fig:k_estim}]{\includegraphics[width=0.45\textwidth, height=7.5cm]{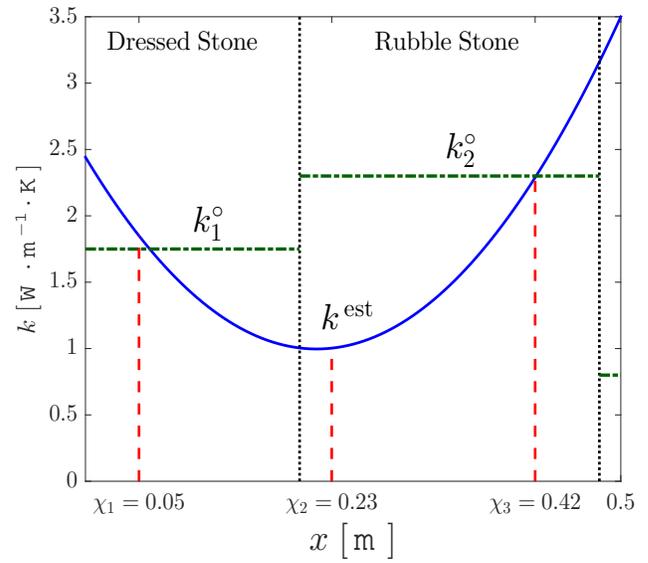}}
\caption{Variation of the computed temperature and the experimental observations for \emph{(a)} first ,\emph{(b)} second and \emph{(c)} third sensor locations; and \emph{(d)} the estimated thermal conductivity representation.}
\label{fig:estim_res}
\end{center}
\end{figure}

Finally, the parameter estimation problem was solved for the quadratic representation of thermal conductivity. The results of three coefficients identification are given in Table~\ref{tab:k_est_all}. 
Figure~\ref{fig:estim_res} shows a comparison between the experimental observations and the numerical results; there was a satisfactory agreement for all points of observations. The computed temperature values were inside the uncertainty boundaries.    
Figure~\ref{fig:k_estim} illustrates how the thermal conductivity changed according to wall depth and the standard values of the thermal conductivity. The thermal property reached its higher values closer to the wall borders. As mentioned in Section~\ref{sec:model_properties}, the profile of the thermal conductivity in Figure~\ref{fig:k_estim} can be justified by the so-called moisture buffering effect~\cite{Carsten_2006}. Indeed, due to the climatic and indoor variations of temperature and relative humidity, heat and mass transfer occurs in the porous wall. The water migration goes from the borders, where the forcing conditions occur, to the centre of the wall. The wall being very thick, the moisture content penetrates only the first centimetres. It is probably higher on the wall borders and lower in its centre. Since the thermal conductivity has a linear relation with the water content in the wall, the profile of thermal conductivity should be similar to the variation of water content in the wall.
   
\begin{figure}[h!]
\begin{center}
\subfigure[$t_{\,\mathrm{ini}} \egal 18/01$ , $x \egal \chi_{\,1}$\label{fig:eps_dd_step}]{\includegraphics[width=0.45\textwidth, height=7.5cm]{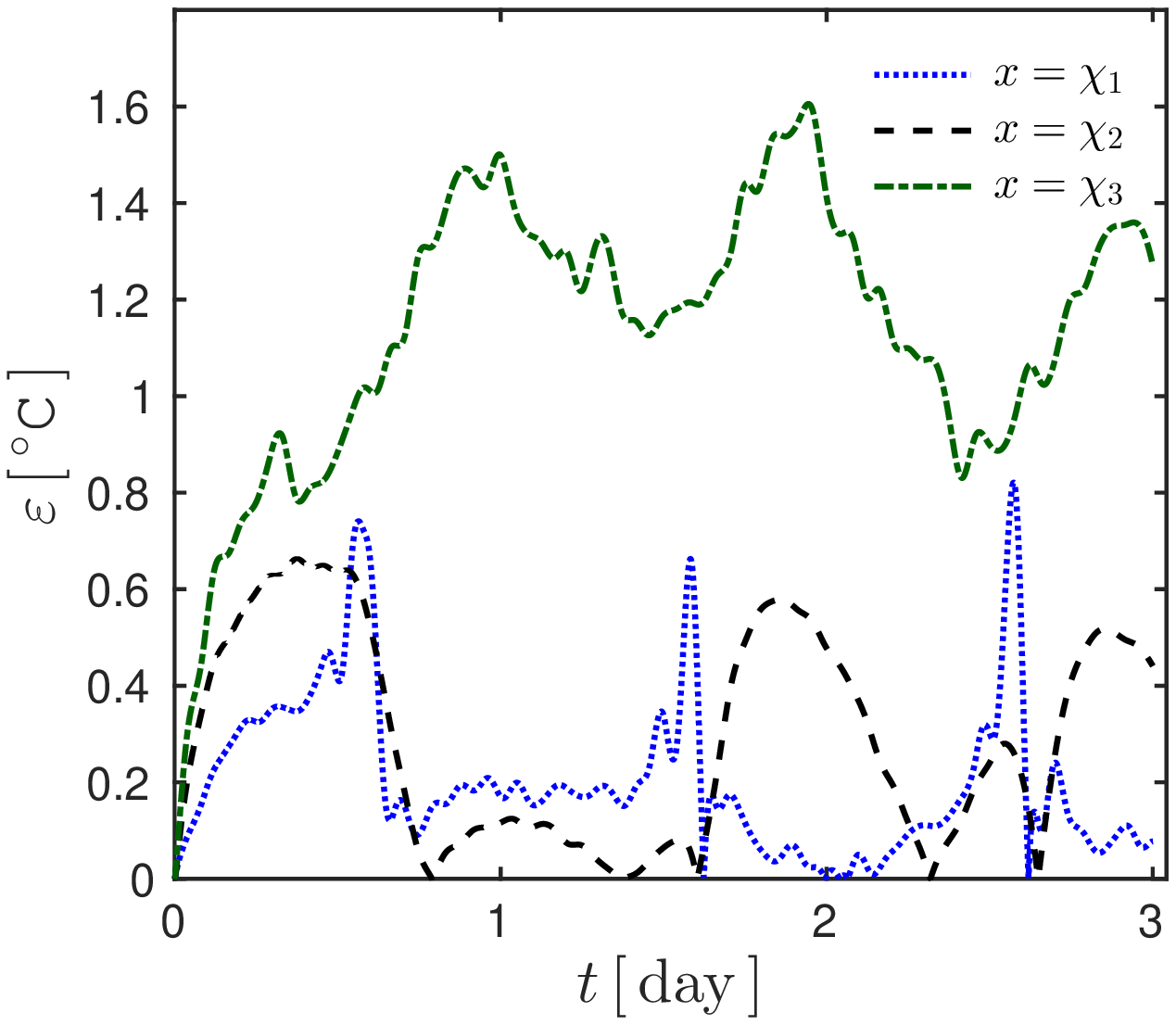}} \quad
\subfigure[$t_{\,\mathrm{ini}} \egal 18/01$ , $x \egal \chi_{\,2}$\label{fig:eps_dd_lin}]{\includegraphics[width=0.45\textwidth, height=7.6cm]{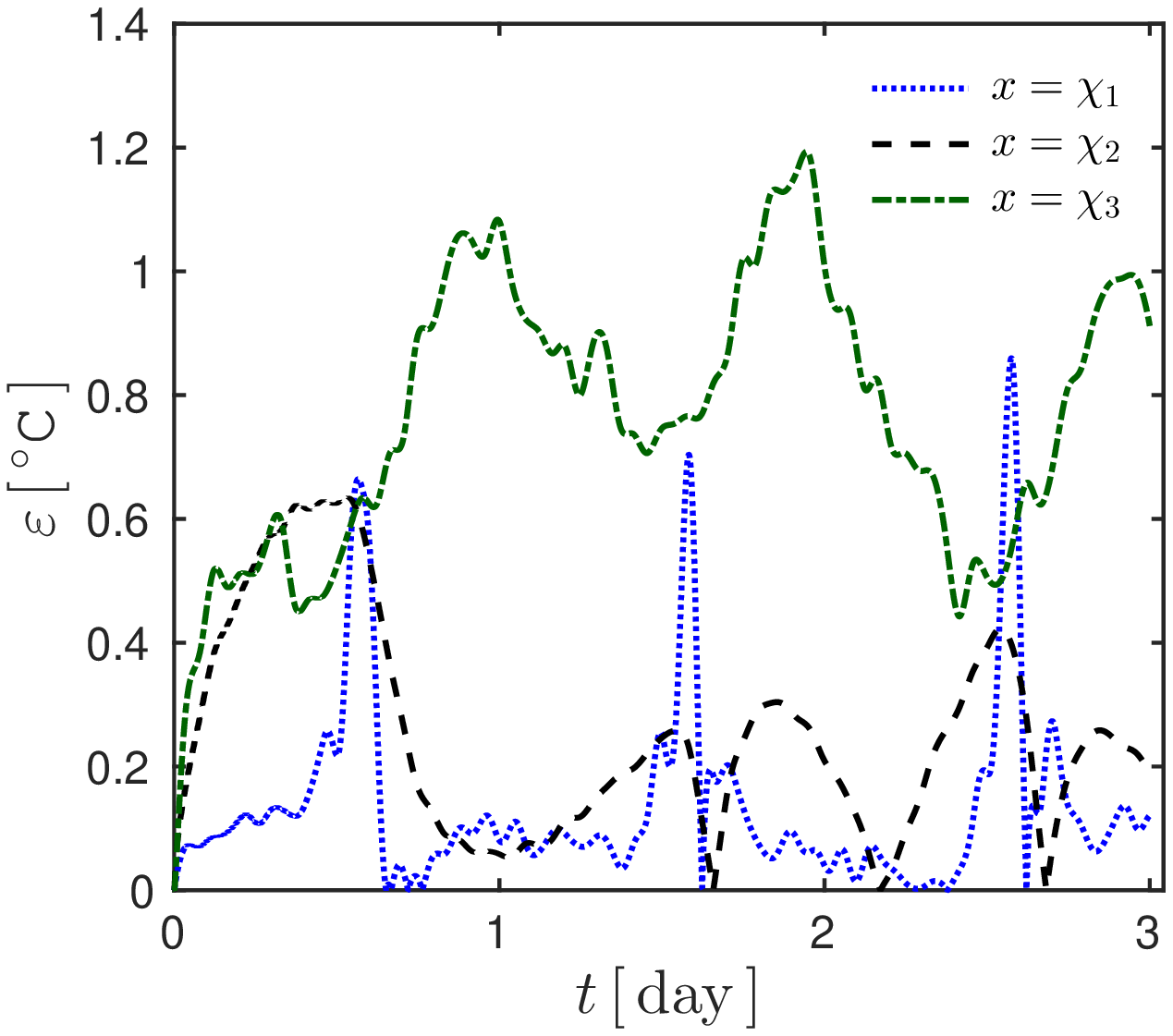}} \quad
\subfigure[$t_{\,\mathrm{ini}} \egal 18/01$ , $x \egal \chi_{\,3}$\label{fig:eps_dd_quad}]{\includegraphics[width=0.45\textwidth, height=7.5cm]{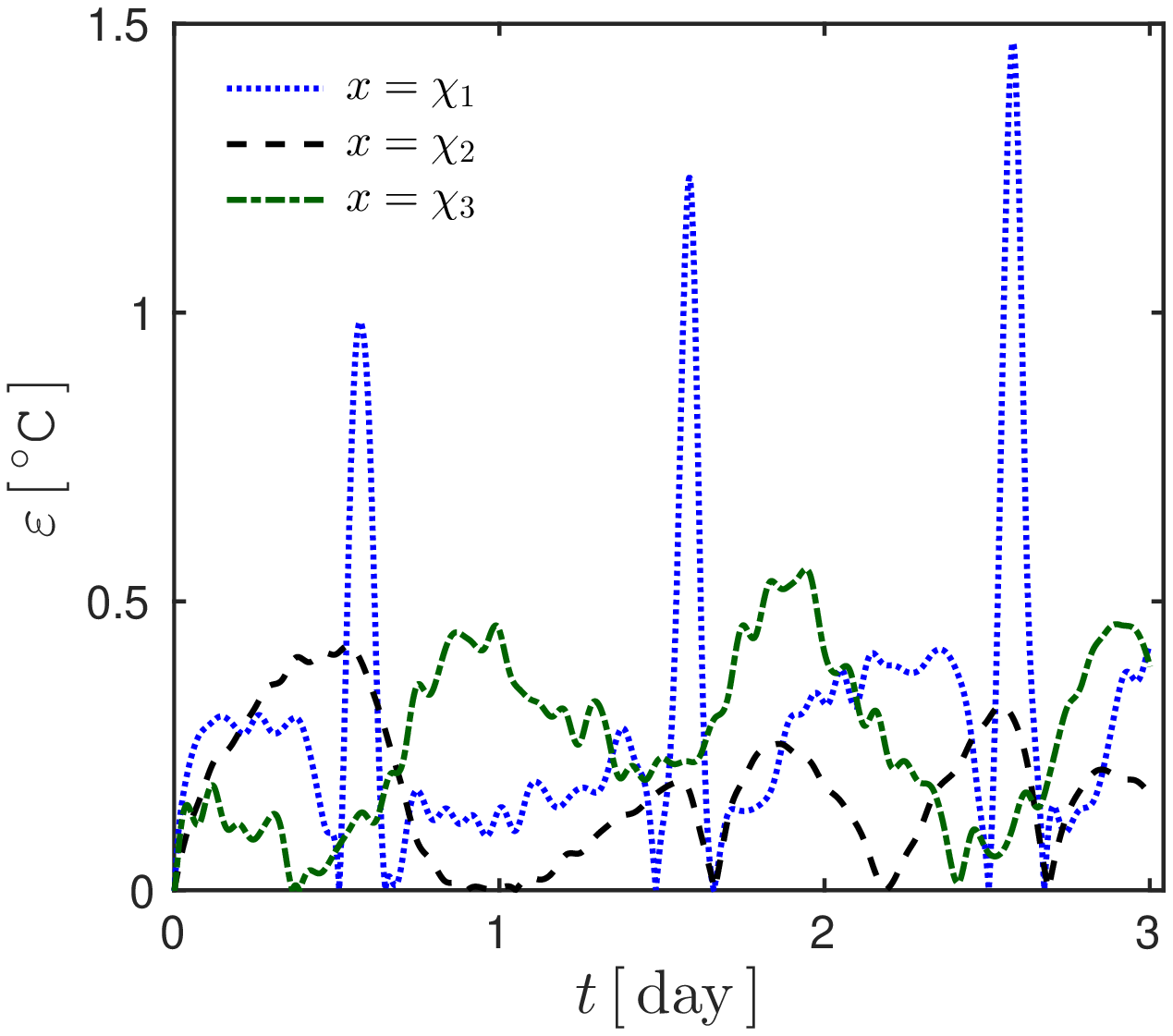}} 
\caption{ Variation of the error $\varepsilon$ of the \emph{(a)} piecewise, \emph{(b)} linear, and \emph{(c)} quadratic representation.}
\label{fig:eps2_dd}
\end{center}
\end{figure}

The residuals for each thermal conductivity representation during the studied period  $t_{\,\mathrm{ini}} \egal 18/01\,,\ \delta\,\tau \egal 3$ can be compared. As shown in Figure~\ref{fig:eps2_dd} the residuals were not correlated regardless of the parameterization, although an error was higher for the piecewise and linear interpolation on the third sensor location. The application of the quadratic interpolation helped to significantly reduce this error. It can be concluded that the quadratic representation of thermal conductivity gave better results than the other two. 

\subsection{Convergence of the optimization process}

\begin{figure}[!ht] 
\centering
\includegraphics[width=0.5\linewidth]{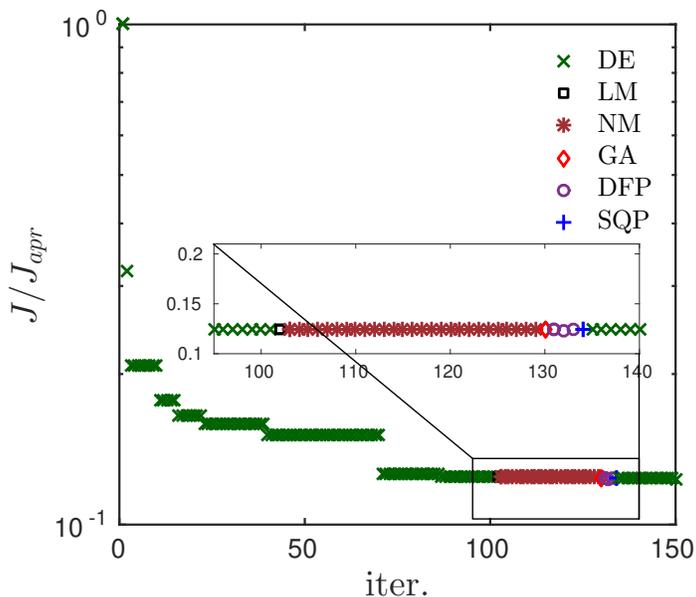}
\caption{Convergence of the optimization process: evolution of the cost function and switch between the methods of the hybrid optimizer.}
\label{fig:conv_hist}
\end{figure}

Figure~\ref{fig:conv_hist} displays the relative variation of cost function according to iterations from the initial point until the convergence criteria.  The process starts with the Differential Evolution (DE) algorithm, and around the hundredth iteration a change of the optimization algorithms occurred, it successively changed from quasi-Newton algorithm of Pshenichny--Danilin (LM) to Nelder--Mead (NM) simplex algorithm, after it shifted to Genetic Algorithm (GA) to Davidon--Fletcher--Powell(DFP) gradient--based algorithm and Sequential Quadratic Programming (SQP). 

The benefit of the reduced measurement plan can be calculated. To obtain an accurate parameter estimation $150$ iterations were required. In addition, the computational time of the numerical model for three days was $4$ seconds, therefore, the overall estimation lasted $10$ minutes. Similarly, as the computational time for one month was $40$ seconds, at least $100$ minutes are necessary to obtain the estimated values. Finally, whole year simulation requires $400$ seconds, and the parameter estimation problem is solved in $17$ hours.   

These evaluations assume that the convergence criteria of the optimization algorithm does not change with the length of observation sequence. Nevertheless, it underlines the advantage of the approach in terms of CPU time. 

\subsection{Evaluating the reliability of the mathematical model.}
\begin{figure}[!ht] 
\centering
\subfigure[\label{fig:quad_dd_x1}]{\includegraphics[width=0.3\textwidth, height=7.4cm]{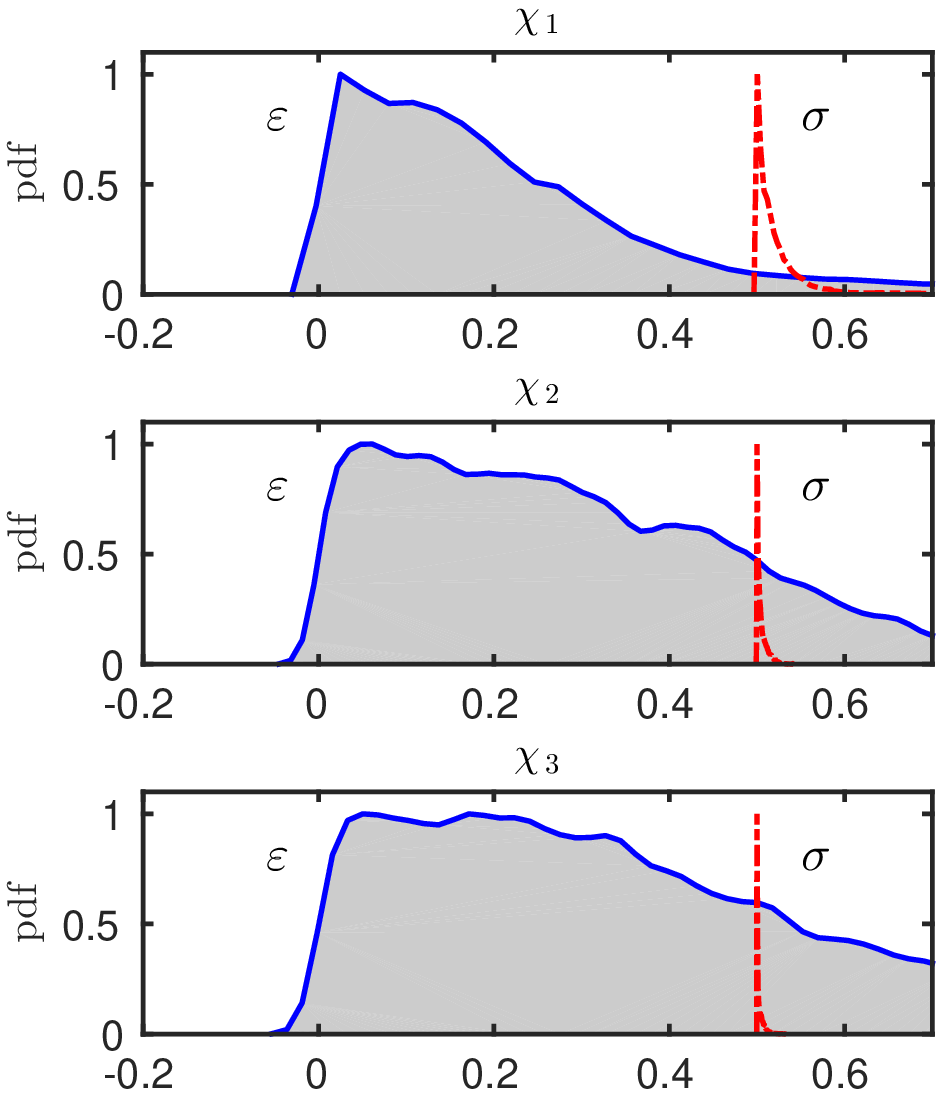}} \qquad
\subfigure[\label{fig:quad_dd_x2}]{\includegraphics[width=0.3\textwidth, height=7.5cm]{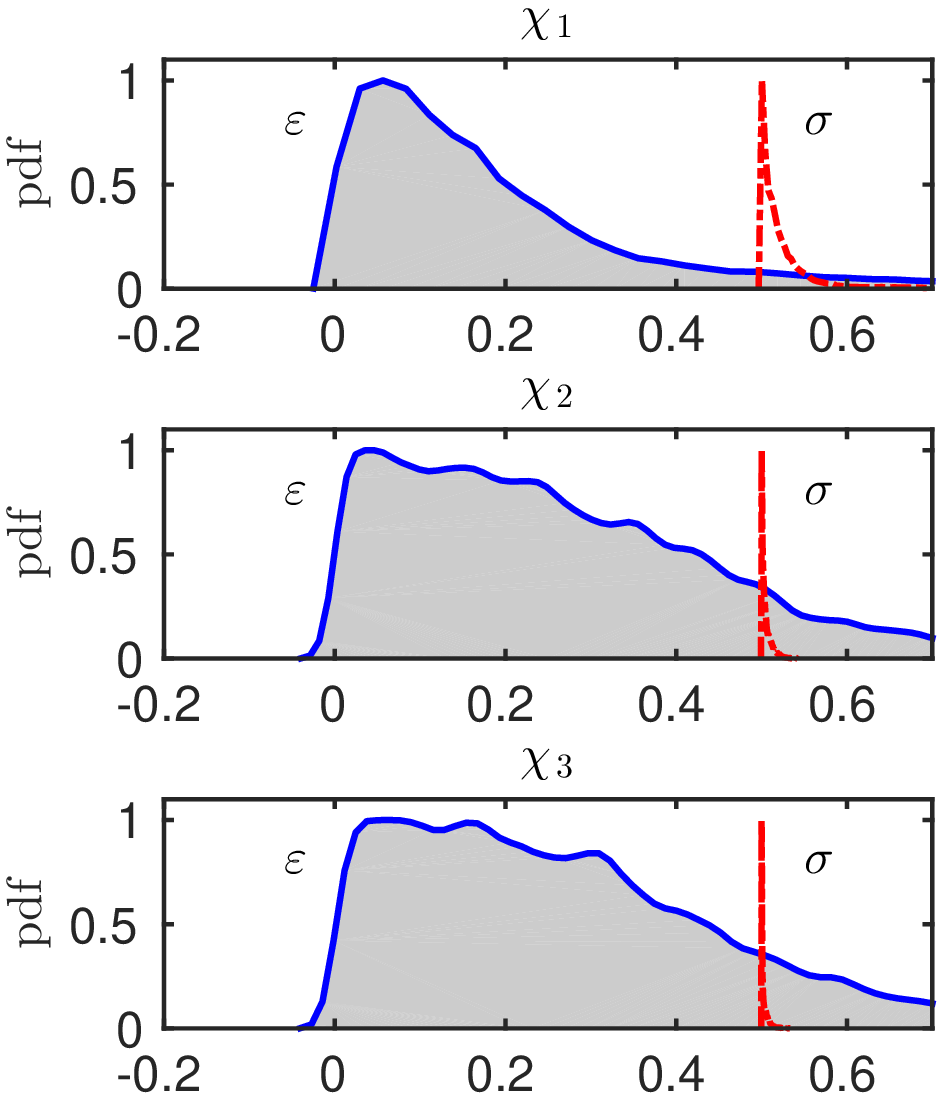}} \qquad
\subfigure[\label{fig:quad_dd_x3}]{\includegraphics[width=0.3\textwidth, height=7.5cm]{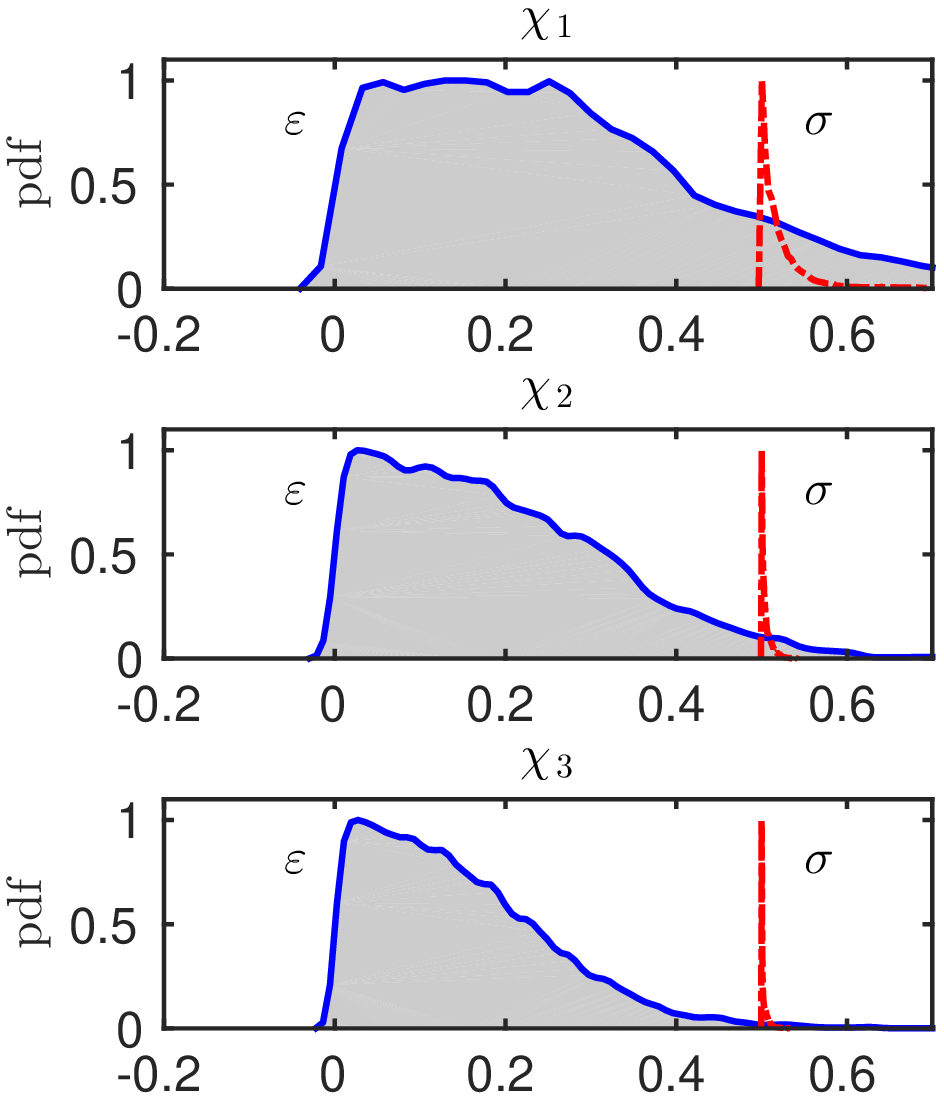}}
\caption{{Normalized probability density function of the error distribution for the piecewise \emph{(a)}, linear \emph{(b)}, and quadratic \emph{(c)} representations compared to the distribution of the measurement uncertainty $\sigma$ values during one year for each sensor location.}}
\label{fig:eps_dens}
\end{figure}

The aim of this section is to investigate the prediction of the model with the experimental data during the whole observation period. This issue is important since the parameter estimation problem has been solved for a reduced measurement plan of three days. Thus, its reliability needs to be evaluated. 
The numerical model was calculated for one year of observations using the estimated parameters.
Figure~\ref{fig:eps_dens} displays the probability density function of the residual for each thermal conductivity representation. In addition, the observation uncertainties for the sensor locations are represented. It can be noted that the mean values of the residuals lay between $0.0\,^{\circ}\,\mathsf{C}$ and $0.5\,^{\circ}\,\mathsf{C}$. Moreover, the computed residuals mean values are smaller than the observations mean values, indicating the satisfactory reliability of the numerical model.

Furthermore, one may conclude that during the one year observations the quadratic representation gave better results on the second and third sensor locations. However, the piecewise and the linear interpolations have better results for the first sensor location. 

\begin{figure}[h!]
\begin{center}
\subfigure[\label{fig:quad_dd_x1}]{\includegraphics[width=0.45\textwidth, height=7.4cm]{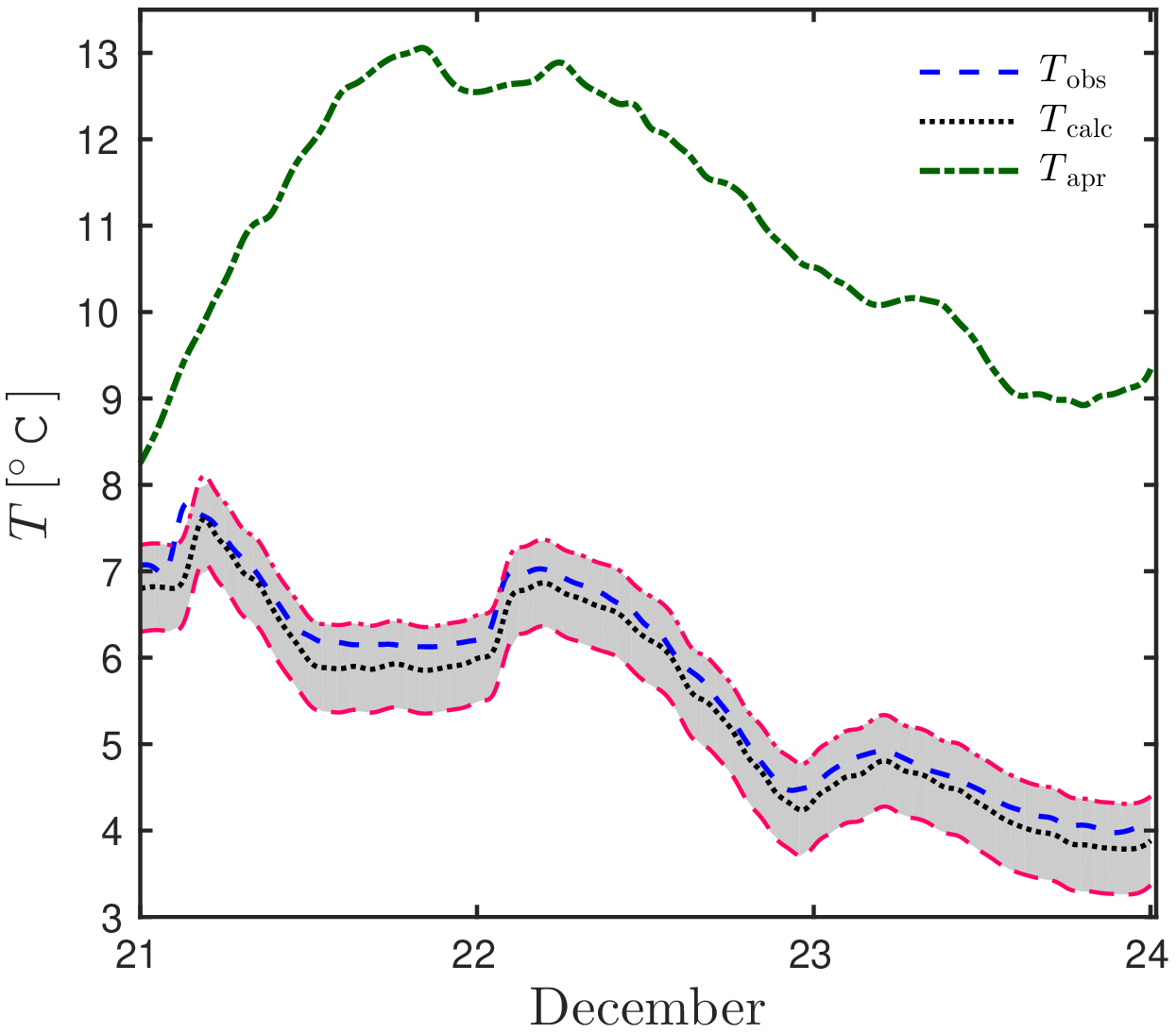}} \quad
\subfigure[\label{fig:quad_dd_x2}]{\includegraphics[width=0.45\textwidth, height=7.5cm]{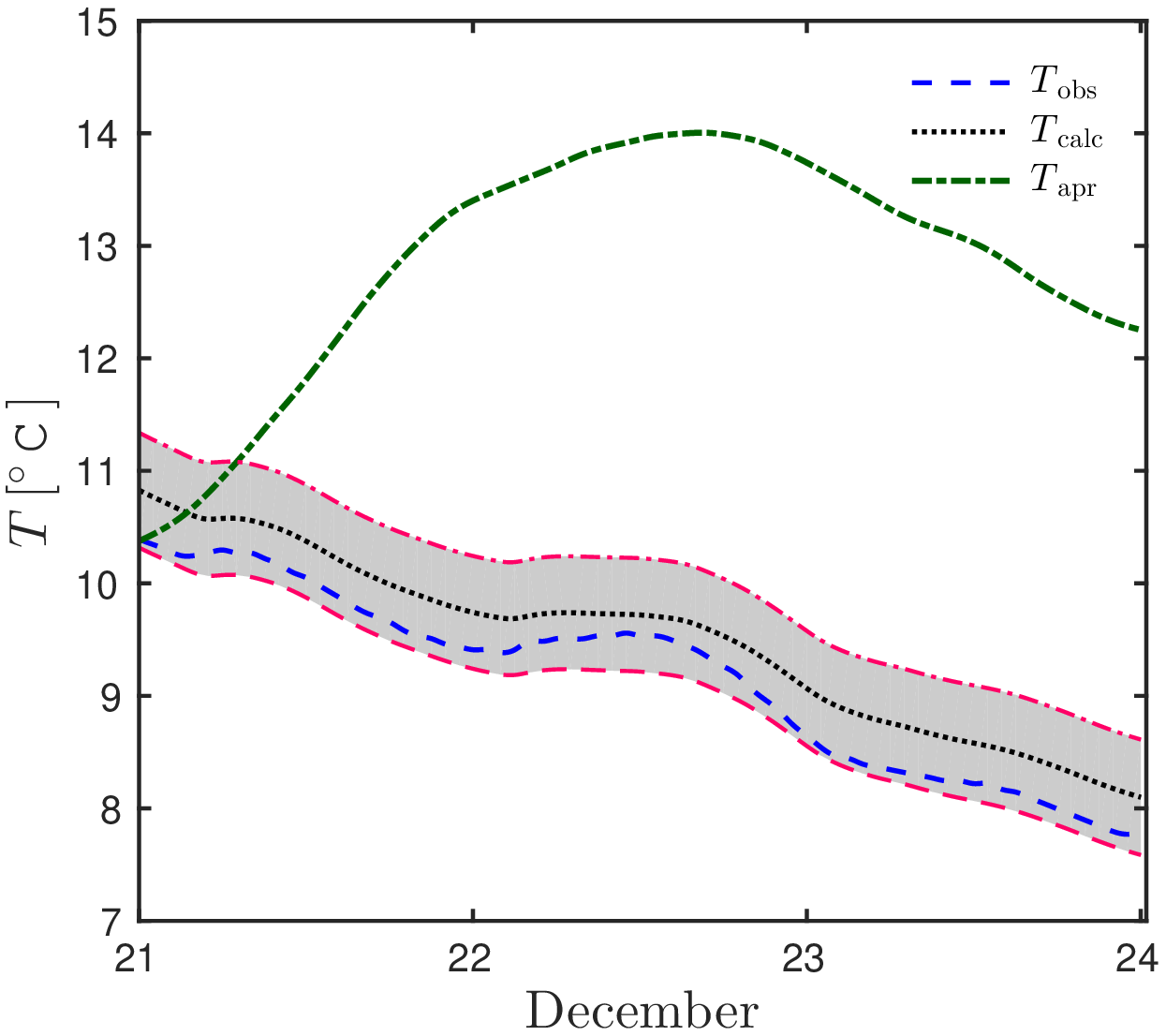}} \quad
\subfigure[\label{fig:quad_dd_x3}]{\includegraphics[width=0.45\textwidth, height=7.5cm]{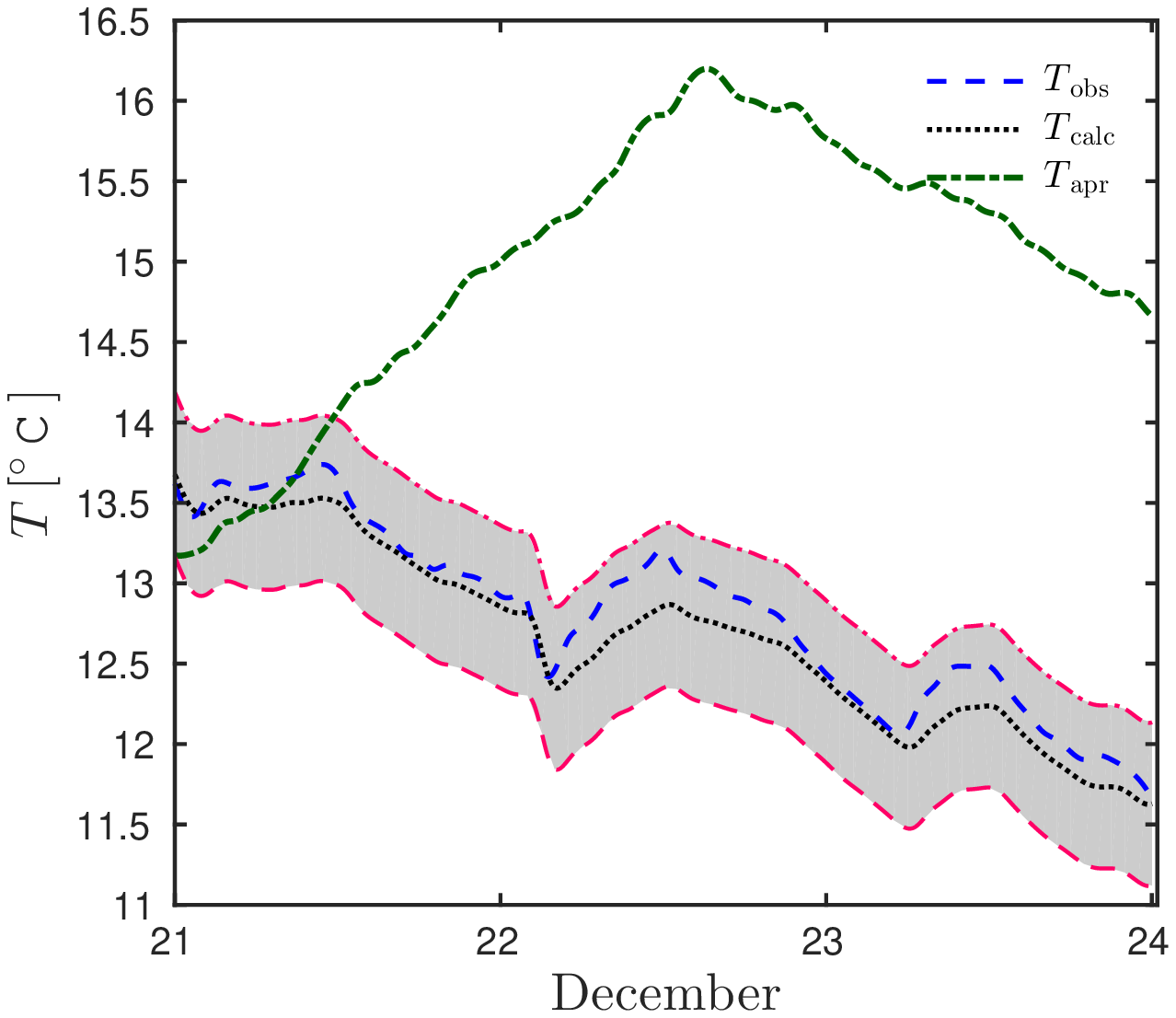}} 
\caption{Variation of the \emph{a priori} temperature, the computed temperature, and the experimental observations during the last three days of observations for \emph{(a)} first ,\emph{(b)} second, and \emph{(c)} third sensor locations.}
\label{fig:last_day}
\end{center}
\end{figure}

Finally, the temperature distribution in the wall was computed during the last three days of the experimental campaign. The calculation was performed using the estimated parameters of the quadratic representation. Figure~\ref{fig:last_day} shows how the predicted temperature varies according to time inside the three sensor locations. A good agreement was obtained between the estimated values and the experimental data; the observations were inside the uncertainty bound of the calculated temperature. Furthermore, the \emph{a priori} information was not reliable for the model prediction.

\section{Conclusion}

In the context of identification of thermophysical parameters of a building wall, the length of observations plays an important role. The accuracy of estimated parameters proportionally depends on the quantity of observations points, however, a long experiment is a computationally expensive task. To find a balance between these two opposite conditions a methodology for determining the optimal duration of the experiment was investigated. The concept of the optimal experiment design (OED) was explored to decrease the duration of an experiment since this approach provides the best accuracy of estimated parameters. The search of the OED was done using the \textsc{Fisher} information matrix, quantifying the amount of information contained in the observations. 

This article studies the wall of a historical building composed of three layers. It was monitored during one year by five sensors: two were placed on the wall's surfaces, while three others within the wall. The estimation of the thermal conductivity of the wall took several steps. First, different spatial parametrizations were proposed to reproduce the variation of thermal conductivity according to the length of the wall. However, only three parametrizations (piecewise, linear and quadratic) were structurally identifiable. Then, the sensitivity coefficients were computed and the practical identifiability of the parameters was discussed. Next, in order to increase the accuracy of the estimation of unknown parameters the D-optimum criterion was analyzed for three measurement plans: one, three and seven days.  It can be noticed that the criterion reaches the maximum value in January, while the minimum is in the summer period.  Then, the three days in January were chosen as an optimal measurement plan as a compromise between the accuracy of the identification method and overall experiment cost. Third, the estimation of thermal conductivity using measurements during these three days was performed by the hybrid optimization method. Results have shown that the implementation of the quadratic representation of thermal conductivity's spatial variation has the best agreement with observations among others two. It can be remarked that a large computation gain is achieved by decreasing the length of observations from one year to three days. Last, the reliability of the model is evaluated by comparing the numerical prediction using the estimated parameters values to the whole year of observations. A very satisfactory agreement is observed highlighting the good reliability of the proposed approach. It indicates that the estimated parameters values using the obtained optimal sequence of three days can be applied to accurately predict a physical phenomenon. 

This study shows that the substantial reduction of the length of observations using OED, nevertheless, provides sufficient information to accurately simulate physical processes. Moreover, the OED approach can be applied for \emph{in situ} measurements only and is not depending on the number of sensors inside the wall. However, it requires \emph{a priori} information of parameters values and depends on a chosen physical model. Further works should focus on the extension of the methodology for more complex mathematical models including for instance coupled heat and mass transfer in porous materials.

\section{Acknowledgments}

The authors acknowledge the Junior Chair Research program “Building performance assessment, evaluation and enhancement” from the University of Savoie Mont Blanc in collaboration with The French Atomic and Alternative Energy Center (CEA) and Scientific and Technical Center for Buildings (CSTB). The authors thanks the grants from the Carnot Institute “Energies du Futur” through the project MN4BAT. The authors thank the IDEX for funding A. Jumabekova travel grant to Florida International University, Miami, USA.

\clearpage
\section*{Nomenclature}

\begin{tabular}{|ccc|}
\hline
\multicolumn{3}{|c|}{\emph{Latin letters}} \\ 
      $c$   &  material volumetric heat capacity     & $\mathsf{[\,J\cdot m^{-3} \cdot K^{-1}\,]}$       \\
      $E$& thermal loads& $\mathsf{[\,W\cdot s \cdot m^{-2} \,]}$      \\
      $h$ & convective heat transfer coefficient & $\mathsf{[\,W \cdot m^{-2} \cdot K^{-1}\,]}$   \\ 
      $j$ & heat flux & $\mathsf{[\,W \cdot m^{-2} \,]}$ \\
      $k$ & thermal conductivity & $\mathsf{[\,W\cdot m^{-1} \cdot K^{-1}\,]}$ \\ 
      $L$ & wall length & $\mathsf{[\,m\,]}$ \\
      $q_{\,\infty}$ & total incident radiation & $\mathsf{[\,W\cdot m^{-2} \,]}$\\
      $T$ & temperature & $\mathsf{[\, K \,]}$\\
      $t$ & time & $\mathsf{[\,s\,]}$\\
      $t_{\,\mathrm{ini}}$ & starting date & $\mathsf{[\,day(s)\,]}$\\
      $\tau_{\,\mathrm{max}}$ & ending date & $\mathsf{[\,day(s)\,]}$\\
      $\delta\,\tau$ & measurement plan & $\mathsf{[\,day(s)\,]}$\\
      $x$ & thickness coordinate direction & $\mathsf{[\,m\,]}$
      \\ \hline
\end{tabular}

\begin{tabular}{|ccc|}
\hline
\multicolumn{3}{|c|}{\emph{Dimensionless values}} \\ 
     $\mathrm{F}$  &  \textsc{Fisher} matrix   & $\mathsf{[\,-\,]}$       \\
     $\mathrm{Fo}$  &  \textsc{Fourier} number   & $\mathsf{[\,-\,]}$      \\
     $\pi$  &  measurement plan   & $\mathsf{[\,-\,]}$       \\
     $\Psi$  &  objective function  & $\mathsf{[\,-\,]}$       \\
      $u$ & temperature field & $\mathsf{[\,-\,]}$ 
      
      \\ \hline
\end{tabular}

\begin{tabular}{|cc|}
\hline
\multicolumn{2}{|c|}{\emph{Subscripts and superscripts}} \\ 
     $L$  &  Left boundary $x\egal 0$        \\
     $R$  &  Right boundary $x\egal L$     \\
    $\star$  &  dimensionless parameter    \\ 
      $\circ$ & \emph{a priori} parameter value  \\
      $\mathrm{est}$ & estimated parameter value     
      \\ \hline
\end{tabular}
\newpage
 
\bibliographystyle{unsrt}
\bibliography{references}

\begin{thebibliography}{10}

\bibitem{IEA_2019}
J.~Dulac, T.~Abergel, and C.~Delmastro.
\newblock Tracking buildings.
\newblock Technical report, International Energy Agency, Paris, 2019.

\bibitem{Sunikka_2012}
M.~Sunikka-Blank and R.~Galvin.
\newblock Introducing the prebound effect: the gap between performance and
  actual energy consumption.
\newblock {\em Building Research \& Information}, 40(3):260--273, 2012.

\bibitem{Wingfield_2008}
J.~Wingfield, M.~Bell, D.~Miles-Shenton, T.~South, and R.J. Lowe.
\newblock Lessons from stamford brook: understanding the gap between designed
  and real performance.
\newblock {\em Final Report of the Partners in Innovation Project CI 39/3/663:
  Evaluating the Impact of an Enhanced Energy Performance Standard on
  Load-Bearing Masonry Domestic Construction, Leeds Metropolitan University},
  2008.

\bibitem{RODLER_2019}
A.~Rodler, S.~Guernouti, and M.~Musy.
\newblock Bayesian inference method for in situ thermal conductivity and heat
  capacity identification: Comparison to iso standard.
\newblock {\em Construction and Building Materials}, 196:574 -- 593, 2019.

\bibitem{BERGER_2017_OED}
J.~Berger, D.~Dutykh, and N.~Mendes.
\newblock On the optimal experiment design for heat and moisture parameter
  estimation.
\newblock {\em Experimental Thermal and Fluid Science}, 81:109 -- 122, 2017.

\bibitem{Berger_2019}
J.~Berger, T.~Busser, D.~Dutykh, and N.~Mendes.
\newblock An efficient method to estimate sorption isotherm curve coefficients.
\newblock {\em Inverse Problems in Science and Engineering}, 27(6):735--772,
  2019.

\bibitem{DALESSANDRO_2019}
G.~D'Alessandro and F.~de~Monte.
\newblock Optimal experiment design for thermal property estimation using a
  boundary condition of the fourth kind with a time-limited heating period.
\newblock {\em International Journal of Heat and Mass Transfer}, 134:1268 --
  1282, 2019.

\bibitem{BERGER_2016}
J.~Berger, H.~R.B. Orlande, N.~Mendes, and S.~Guernouti.
\newblock Bayesian inference for estimating thermal properties of a historic
  building wall.
\newblock {\em Building and Environment}, 106:327 -- 339, 2016.

\bibitem{beck_1977}
J.V. Beck and K.J. Arnold.
\newblock {\em Parameter estimation in engineering and science}.
\newblock Wiley, 1977.

\bibitem{stephenson1971}
D.~G. Stephenson and G.P. Mitalas.
\newblock Calculation of heat conduction transfer functions for multi-layers
  slabs.
\newblock {\em Air Cond. Engrs. Trans;(United States)}, 77, 1971.

\bibitem{Ozisik}
M.N. Ozisik and H.R.B. Orlande.
\newblock {\em Inverse Heat Transfer - Fundamentals and Applications}.
\newblock CRC Press, New York, 2000.

\bibitem{Rode_2004}
C.~Rode, N.~Mendes, K.~Grau, and I.~Walker.
\newblock Evaluation of moisture buffer effects by performing whole-building
  simulations.
\newblock {\em ASHRAE Transactions}, 110:783--794, 01 2004.

\bibitem{Du_Fort_1953}
E.~C.~Du Fort and S.~P. Frankel.
\newblock Stability conditions in the numerical treatment of parabolic
  differential equations.
\newblock {\em Mathematical Tables and Other Aids to Computation},
  7(43):135--152, 1953.

\bibitem{Gasparin_2017a}
S.~Gasparin, J.~Berger, D.~Dutykh, and N.~Mendes.
\newblock Stable explicit schemes for simulation of nonlinear moisture transfer
  in porous materials.
\newblock {\em Journal of Building Performance Simulation}, 11(2):129--144,
  2018.

\bibitem{Gasparin_2017b}
S.~Gasparin, J.~Berger, D.~Dutykh, and N.~Mendes.
\newblock An improved explicit scheme for whole-building hygrothermal
  simulation.
\newblock {\em Building Simulation}, 11(3):465--481, 2018.

\bibitem{Walter_1982}
E.~Walter and Y.~Lecourtier.
\newblock Global approaches to identifiability testing for linear and nonlinear
  state space models.
\newblock {\em Mathematics and Computers in Simulation}, 24(6):472--482, 1982.

\bibitem{Finsterle_2015}
S.~Finsterle.
\newblock Practical notes on local data-worth analysis.
\newblock {\em Water Resources Research}, 51(12):9904–9924, 2015.

\bibitem{Walter_1990}
E.~Walter and L.~Pronzato.
\newblock Qualitative and quantitative experiment design for phenomenological
  models; a survey.
\newblock {\em Automatica}, 26(2):195--213, 1990.

\bibitem{Ucinski}
D.~Ucinski.
\newblock {\em Optimal Measurement Methods for Distributed Parameter System
  Identification}.
\newblock CRC Press, New York, 2004.

\bibitem{Dulikravich_1999}
G.~S. Dulikravich, T.~J. Martin, B.~H. Dennis, and N.~F. Foster.
\newblock {\em Multidisciplinary hybrid constrained GA optimization},
  chapter~12, pages 231--260.
\newblock John Wiley and Sons, Finland, 1999.

\bibitem{Dulikravich_2013}
G.~S. Dulikravich, T.~J. Martin, M.~J. Colaco, and E.~Inclan.
\newblock Automatic switching algorithms in hybrid single-objective
  optimization.
\newblock {\em FME Transactions}, 41:167--179, 01 2013.

\bibitem{Davidon_1959}
W.C. Davidon.
\newblock Variable metric method for minimization.
\newblock {\em ANL-5990}, 5 1959.

\bibitem{Fletcher_1963}
R.~Fletcher and M.~J.~D. Powell.
\newblock {A Rapidly Convergent Descent Method for Minimization}.
\newblock {\em The Computer Journal}, 6(2):163--168, 08 1963.

\bibitem{Goldberg_1989}
D.~E. Goldberg.
\newblock {\em Genetic Algorithms in Search, Optimization and Machine
  Learning}.
\newblock Addison-Wesley Longman Publishing Co., Inc., USA, 1st edition, 1989.

\bibitem{Nelder_1965}
J.~A. Nelder and R.~Mead.
\newblock {A Simplex Method for Function Minimization}.
\newblock {\em The Computer Journal}, 7(4):308--313, 01 1965.

\bibitem{Storn_1996}
R.~{Storn} and K.~{Price}.
\newblock Minimizing the real functions of the icec'96 contest by differential
  evolution.
\newblock In {\em Proceedings of IEEE International Conference on Evolutionary
  Computation}, pages 842--844, May 1996.

\bibitem{SQP_1999}
A.~Tits, J.~Zhou, and C.~Lawrence.
\newblock User's guide for ffsqp version 3.7: A fortran code for solving
  constrained nonlinear (minimax) optimization problems, generating iterates
  satisfying all inequality and linear constraints.
\newblock 02 1999.

\bibitem{LM_1969}
B.N. Pshenichny and Y.M. Danilin.
\newblock {\em Numerical Methods in Extremal Problems}.
\newblock MIR Publishers, Moscow, 1969.

\bibitem{fr_std}
Journal~Officiel de~la République~Française.
\newblock Arrêté du 26 octobre 2010 relatif aux caractéristiques thermiques
  et aux exigences de performance énergétique des bâtiments nouveaux et des
  parties nouvelles de bâtiments, 2010.

\bibitem{Carsten_2006}
C.~Rode and R.H. Peuhkur.
\newblock {\em The Concept of Moisture Buffer Value of Building Materials and
  its Application in Building Design}, volume III, pages 57--62.
\newblock 2006.

\end{thebibliography}
\end{document}